%% file: paper.tex
\documentclass[11pt,letterpaper]{article}

\usepackage{amsmath,amsthm,amsfonts,amssymb}
\usepackage{thm-restate}
\usepackage{fullpage}
\usepackage[utf8]{inputenc}
\usepackage[dvipsnames]{xcolor}
\usepackage{xspace,enumerate}
\usepackage[shortlabels]{enumitem}
\usepackage[hypertexnames=false,colorlinks=true,urlcolor=Blue,citecolor=Green,linkcolor=BrickRed]{hyperref}
\usepackage[capitalise]{cleveref}
\usepackage[OT4]{fontenc}
\usepackage[ruled]{algorithm}
\usepackage{ifpdf}
\usepackage{todonotes}
\usepackage{enumerate}
\usepackage{thmtools}
\usepackage{mathtools}
\usepackage[noend]{algpseudocode}
\usepackage{textpos}

\usepackage[margin=1in]{geometry}

\newif\ifshort
\newif\iffull
\fulltrue



\theoremstyle{plain}
\newtheorem{theorem}{Theorem}[section]
\newtheorem{lemma}[theorem]{Lemma}
\newtheorem{corollary}[theorem]{Corollary}

\newtheorem{fact}[theorem]{Fact}
\newtheorem{observation}[theorem]{Observation}
\newtheorem{Definition}[theorem]{Definition}

\newtheorem{claim}[theorem]{Claim}

\def\poly{\operatorname{poly}}
\def\polylog{\operatorname{polylog}}

\newcommand{\email}[1]{\href{mailto:#1}{#1}}

\title{Exact Shortest Paths with Rational Weights on the Word RAM\thanks{This work is a part of project BOBR (WN, MS) that has received funding from the European Research Council (ERC) under the European Union’s Horizon 2020 research and innovation programme (grant agreement No. 948057). Adam Karczmarz was partially supported by the ERC CoG grant TUgbOAT no 772346 and the National Science Centre (NCN) grant no. 2022/47/D/ST6/02184.}}

\date{}
\author{Adam Karczmarz\thanks{University of Warsaw and IDEAS NCBR, Poland. \email{a.karczmarz@mimuw.edu.pl}.}
\and Wojciech Nadara\thanks{University of Warsaw, Poland. \email{w.nadara@mimuw.edu.pl}.}
\and Marek Sokołowski\thanks{University of Warsaw, Poland. \email{marek.sokolowski@mimuw.edu.pl}.}}

\newcommand{\Oh}{\ensuremath{O}}
\newcommand{\Ot}{\ensuremath{\widetilde{O}}}
\newcommand{\eps}{\ensuremath{\epsilon}}
\newcommand{\dist}{\delta}
\newcommand{\len}{w}
\newcommand{\wei}{w}

\newcommand{\ceil}[1]{\left\lceil #1 \right\rceil}

\newcommand{\Q}{\mathbb{Q}}
\newcommand{\R}{\mathbb{R}}
\newcommand{\Qq[1]}{\mathbb{Q}^{(#1)}}
\newcommand{\Z}{\mathbb{Z}}

\begin{document}

\maketitle
\thispagestyle{empty}

\begin{textblock}{20}(-1.8, 9.2)
	\includegraphics[width=40px]{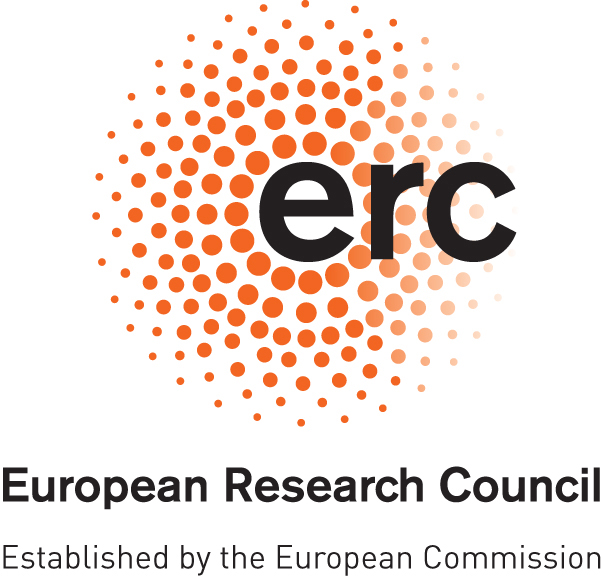}%
\end{textblock}
\begin{textblock}{20}(-2.05, 9.5)
	\includegraphics[width=60px]{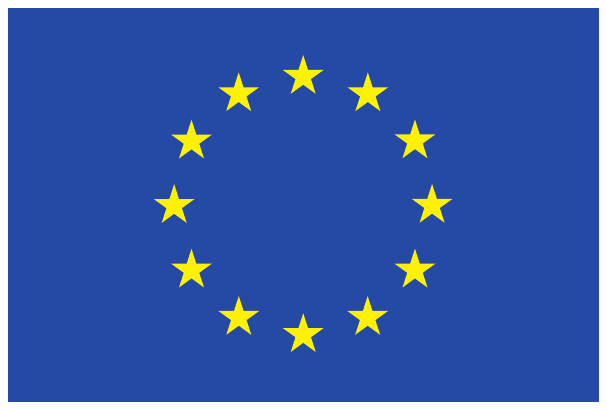}%
\end{textblock}

\begin{abstract}
\input{abstract}

\end{abstract}

\clearpage
\setcounter{page}{1}

\input{intro}

\input{preliminaries}

\input{overview}


\section{The algorithm for the non-negative case}\label{s:sssp-details}

\input{fast-sssp}

\input{scaling}

\section{Negative-SSSP and negative cycle detection in $O(n^{2.5})$}
\input{neg-cycle}

\bibliographystyle{alpha}

\bibliography{references}

\appendix

\input{appendix}

\end{document}

%% file: abstract.tex
  Exact computation of shortest paths in weighted graphs has been
  traditionally studied in one of two settings.
  First, one can assume that the edge weights are real numbers and all the performed
  operations on reals (typically comparisons and additions) take constant time.
  Classical Dijkstra's and Bellman-Ford algorithms have been described in this setting.
  
  More efficient exact shortest paths algorithms have been obtained for integer-weighted graphs.
  Integrality assumption not only enables faster algorithms but
  also allows implementing the aforementioned algorithms in a much more realistic word RAM model
  where only arithmetic operations on $O(\log{n})$-bit integers are performed in constant time.
  
  On the word RAM one can as efficiently exactly encode even
  \emph{rational-weighted} instances with $O(\log{n})$-bit numerators and denominators.
  However, the known exact real-weighted shortest paths algorithms, run on such a rational input, can easily
  encounter intermediate values of $\Theta(n)$ bits if represented exactly.
  This leads to a factor-$\Omega(n)$ slowdown on the word RAM.
  At the same time, the scaling algorithms suited for integer weights do not produce exact
  solutions for rational inputs without dramatically increasing their accuracy.

  In this paper, we design randomized exact single-source shortest paths algorithms for rational-weighted graphs on the word RAM.
  Most importantly, in the non-negative case, we obtain a near-linear time algorithm matching
  Dijkstra's algorithm running time up to polylogarithmic factors.
  In presence
  of negative weights, we give an $\Ot(n^{2.5})$-time algorithm breaking through
  the best known strongly polynomial bound attained by Bellman-Ford for sufficiently dense graphs.

%% file: intro.tex
\section{Introduction}
Fundamental polynomial-time optimization problems on \emph{weighted} graphs, such as computing shortest
paths and maximum flow, have been traditionally studied from the point of view
of \emph{exact} computation in two ways. One way is to assume that the edge weights
are given as real numbers, and the computational model allows performing elementary
operations on reals\footnote{At least comparisons and additions, sometimes also other
basic arithmetic operations.}
in constant time in a black-box manner.
This convenient extension of the popular \emph{word RAM} model with an $O(\log{n})$-bit word~\cite{FredmanW93} is known as the \emph{real RAM}~\cite{shamos1978computational} and is a standard e.g. in computational geometry.
Exact algorithms in this model are typically \emph{strongly polynomial}, that
is, their time cost can be bounded by a function of $n$
and $m$ (the number of vertices and edges, respectively) exclusively.\footnote{Formally, for an algorithm to be strongly polynomial,
one also requires it runs in polynomial
time on (binary-encoded) rational inputs on a standard Turing machine.
This is the case for every result we refer to in this paper.}
Many of the classical graph algorithms have been described in this model: Dijkstra's~\cite{Dijkstra59}
and Bellman-Ford~\cite{bellman1958routing} algorithms for single-source shortest paths, maximum flow algorithms
such as Edmonds-Karp~\cite{EdmondsK72} and push-relabel~\cite{GoldbergT88}, the Hungarian method for the assignment problem~\cite{kuhn1955hungarian}, etc.

While the real RAM model is simple and convenient to work with, 
unless further assumptions are made,
it is unrealistic to assume
that arithmetic operations on the input and intermediate values take constant time.
To circumvent this problem, one possible approach is to use floating point arithmetic. But then we cannot hope for
producing exact solutions.
An alternative way is to restrict the domain.
For the single-source shortest paths (SSSP) problem, assuming that input edge
weights are \emph{integers} fitting in a machine word
allows a realistic encoding of the input and transferring the $O(m+n\log{n})$ and $O(mn)$ bounds of Dijkstra's and Bellman-Ford (resp.) algorithms
to a much more realistic word RAM model.

\paragraph{Integrality and scaling.}
In fact, limiting attention to \emph{integer data} has become a well-established
approach to studying exact optimization problems on graphs,
especially given that integrality allows for tighter bounds.
The non-negative SSSP problem with integer weights bounded by $W$
can be solved in ${O(m+n\log\log{(\min(n,W))})}$ time~\cite{Thorup04}.
It has been also known since the eighties that the integer variant of the negative-weight SSSP problem
can be solved \emph{polynomially faster} then Bellman-Ford
if the maximum absolute weight $W$ is polynomially bounded in~$n$~\cite{Gabow85}.
Gabow's algorithm leverages the \emph{scaling} framwork whose usefulness for
graph algorithms has been shown already by Edmonds and Karp~\cite{EdmondsK72}.
After a long line of combinatorial scaling~\cite{GabowT89, Goldberg95} and interior point method-based \cite{CohenMSV17, AxiotisMV20, BrandLNPSS0W20, ChenKLPGS22} improvements,
Bernstein, Nanongkai, and Wulff-Nilsen discovered
a combinatorial $\Ot(m\log{W})$ time algorithm~\cite{BernsteinNW22} for negatively-weighted SSSP (slightly improved even more recently~\cite{abs-2304-05279})
that falls into the original scaling framework.

Let us now briefly explain how the integer data assumption is exploited in scaling algorithms.
Roughly speaking, the scaling framework (as described in~\cite{GoldbergT90}) proceeds by iteratively computing
more and more accurate solutions. With each iteration (which e.g.\ in~\cite{BernsteinNW22} takes near-linear time), the absolute approximation error is reduced by a constant factor.
As a result, the final running time bound always involves a $\log(1/\delta)$ factor,
where~$\delta$ is the desired relative accuracy parameter, corresponding to the number of iterations made.
For integer data, the accuracy ${\delta=1/\poly(n,W)}$ is enough to enable
efficient rounding of an approximate solution to an exact one.
In principle, one could run a scaling algorithms on \emph{real-valued} (e.g., irrational) data in the stronger real RAM model,
but it could happen that an exact optimum is never reached.
Moreover, since the running time depends on the desired
relative accuracy (and thus on $W$), it cannot be bounded using $n$ and~$m$ exclusively.
This in turn deems the framework inherently \emph{weakly polynomial}.

\paragraph{Rational data.}
\emph{Rational-weighted} graphs, with both numerators and denominators fitting in
a machine word, can be represented on a word RAM machine as exactly and as concisely as integer-weighted instances.
Therefore, it is surprising to note that -- to the best of our knowledge -- the most fundamental polynomial exact optimization problems on graphs
have been thoroughly studied only either in
an idealized but unrealistic real RAM model,
or under the integrality assumption that is not as general as is naturally allowed by the word RAM model.
It is thus interesting to ask:
\begin{quote}
  \centering
  \textit{
  1. Can optimization problems on rational-weighted graphs be solved on the word RAM exactly and as efficiently as on integer-weighted graphs?
  }
\end{quote}
The answer to the question is trivially affirmative for some problems, such as the minimum spanning tree,
where the solution depends only on the linear order on the input weights.
However, for problems such as single-source shortest paths or maximum flow, the answer
is far from obvious.
\paragraph{Case study: SSSP.}
Consider the exact SSSP problem, where the desired output is a shortest paths tree from a given source.
Let us call a rational number \emph{short} if both its numerator and denominator fit in a machine word.
There are two natural ways to construct an SSSP algorithm for rational-weighted
graphs. First, one can simply adapt a strongly polynomial algorithm (Dijkstra's or Bellman-Ford)
by replacing the abstract real arithmetic with an exact rational arithmetic in a black-box way.
It is easy to see that the intermediate values arising when these algorithms
proceed represent lengths of some paths consisting of at most $n$ edges.
Since a sum of $n$ short rational numbers is an $\Ot(n)$-bit rational number
and arithmetic operations on $\leq k$-bit rational numbers take $\Ot(k)$ time on the word RAM,
the rational arithmetic overhead incurred is $\Ot(n)$.
We conclude that Dijkstra's algorithm on rational data can be implemented
in $\Ot(mn)$ time on the word RAM, whereas Bellman-Ford in $\Ot(mn^2)$ time.
This is significantly slower than in the integral case.

Another natural approach is to apply the scaling framework.\footnote{Whereas scaling has been most successfully applied to negatively-weighted
SSSP, there also exists a scaling algorithm for the non-negative SSSP~\cite{Gabow85}.} There, one quickly recognizes
that polynomial accuracy $\delta$ is not sufficient to get an exact answer for rational data. For example,
distinct weights of two length-$n$ paths can differ by $2^{-\widetilde{\Omega}(n)}$ 
\iffull
(as shown in Appendix~\ref{s:small-diff}).
\fi
\ifshort
(as shown in the full version).
\fi
This is as opposed to the integer case, where the minimum possible difference is $1$.
Thus, if the scaling framework is applied out-of-the box, the accuracy
parameter $\delta$ should be set to be at least exponential in~$n$.
Even if one managed to use only $\Ot(1)$-bit intermediate values in the computation, using an extreme-precision version
of a near-linear scaling
SSSP algorithm naively costs at least $\Ot(mn)$ time as well.

For negatively-weighted SSSP there is currently a large $\widetilde{\Omega}(n)$
gap between the best known weakly polynomial bound~\cite{BernsteinNW22} (in the integer case) and the best strongly polynomial bound
of Bellman-Ford.
It is tempting to ask whether this is related to the relatively low precision required
by the integer case compared to the rational case.
This motivates another general question:
\begin{quote}
  \centering
  \textit{
  2. Can one solve optimization problems on rational-weighted graphs on the word RAM exactly beyond the respective best-known strongly
  polynomial bounds on the real RAM?
  }
\end{quote}

\paragraph{Inevitable non-integrality.} A further motivation for studying rational-weighted exact optimization
comes from the fact that for some important graph problems running into rational data is inevitable
even if the input data is as integral as one could imagine.
In some of these cases, exact optimization algorithms have never been analyzed in a fine-grained way in any realistic arithmetic model.
One example is the generalized flow problem, which extends the standard maximum flow problem
with linear arc gains $\gamma(e)$:
if $f$ units of flow enter an edge $e=uv$ at vertex $u$, $\gamma(e)\cdot f$ units of flow
arrive at $v$. The state-of-the-art exact combinatorial algorithm~\cite{Radzik04} is weakly polynomial\footnote{Strongly polynomial algorithms for generalized max-flow have been found only relatively recently~\cite{Vegh17, OlverV20}.}: the edge capacities are assumed to be integral, gains rational, and all the integers
used to encode the input are bounded by $W$. However, the arithmetic model assumed therein is in fact the real RAM.

An even simpler textbook example (e.g.,~\cite{cormen}, also arising in the generalized flow literature, e.g.,~\cite{TardosW98,GoldfarbJO97}) is \emph{arbitrage detection}.
For $n$ currencies, pairwise exchange rates are given as rational numbers.\footnote{Clearly, some of the exchange rates should be positive numbers less than $1$.} The objective
is to check whether one can start with a unit of some currency $a$, do some trading, and end up with more than one unit of $a$,
i.e., make pure profit.
A classic solution to this problem (see, e.g.,~\cite{cormensolutions}) is to reduce to negative cycle detection on
a complete graph of currencies where the weight of an edge $ab$ equals the logarithm of the exchange rate between $a$ and~$b$.
The log-transformation is already problematic in a realistic arithmetic model (and even
in the basic real RAM model) since
a logarithm can be irrational. One can avoid logarithms by directly adapting the Bellman-Ford
algorithm to treat the product of exchange rates on a path as the path weight.
However, even a product of $k$ integers (let alone rational numbers) may require $\Ot(k)$ bits to be represented,
which again introduces a non-trivial arithmetic overhead.

\subsection{Our contribution}
In this paper we initiate the study of the word RAM complexity of optimization on rational-weighted graphs
with the goal of providing answers to the posed questions 1 and 2.
Of course, the sought answers might be problem-specific.
Therefore, in this paper we put the focus on the single-source shortest paths problem, perhaps the most
basic problem on weighted directed graphs where the challenges specific
to rational arithmetic do manifest themselves.

First of all, we show that the non-negative single-source shortest paths problem
can be solved on short rational data with only polylogarithmic slowdown wrt. the integer case and thus a factor-$O(n)$ faster than naively.
We assume a standard word-RAM model with $O(\log{n})$-bit machine
word.

\begin{theorem}\label{t:sssp}
Let $G$ be a directed graph with non-negative rational weights whose numerators
and denominators fit in a machine word. There exists a Monte Carlo randomized 
algorithm computing a single-source shortest path tree in $G$
  from any chosen source vertex in $\Ot(n+m)$ time (w.h.p.\footnote{
    With high probability, that is, with probability at least $1-1/n^\gamma$ for any chosen constant $\gamma>0$.}).
\end{theorem}
Theorem~\ref{t:sssp} provides an affirmative answer to our first motivating question
up to polylogarithmic factors.
Our algorithm also matches the best known near-linear strongly polynomial bound (in the real RAM model) 
wrt. $\Ot(\cdot)$ notation.
A single-source shortest paths tree $T\subseteq G$ from $s$ can be preprocessed in near-linear
time to be able to report any $s\to t$, $t\in V$ distance in $\Ot(n)$ time (see Lemma~\ref{l:tree-query}).
Note that explicitly outputting all the $n$ distances from $s$ might require
$\Omega(n^2)$ time since one can easily construct an instance where $\Theta(n)$ distances
require $\Theta(n)$ bits each.

We also study the more general variant of the SSSP problem allowing negative weights.
First, we show
\iffull
(Lemma~\ref{l:scaling})
\fi
\ifshort
(in the full version)
\fi
that the scaling framework can be applied for rational-weighted SSSP
in such a way that every iteration boils down\footnote{This is non-obvious since the known scaling algorithms,
tailored to handle the integral case only, do not have to non-trivially control the bit-length of the intermediate values.
The values encountered for polynomial accuracy required in the integer case clearly use only $O(\polylog{n})$ bits.}
to solving an instance with polynomial
integer weights which costs $\Ot(m)$ time~\cite{BernsteinNW22, abs-2304-05279}.
Consequently, and since exponential accuracy ${\delta=O(n^{-n})}$ is sufficient,
negatively-weighted SSSP can be solved in $\Ot(nm)$ time for rational weights with numerators
and denominators fitting in an $O(\log{n})$-bit machine word. This again matches
the problem's best known strongly polynomial bound without ever breaking through it, though.

By combining the scaling framework (used for moderate accuracy of roughly $n^{-\sqrt{n}}$) 
with data structures developed for obtaining Theorem~\ref{t:sssp} and some further
insights, we obtain the following.
\begin{theorem}\label{t:negsssp}
Let $G$ be a digraph with rational weights whose numerators and denominators fit
  in a machine word. There exists a Monte Carlo
  \iffull
  \fi
  randomized algorithm computing
  a single-source shortest paths tree in $G$ from a chosen source
  (or detecting a negative cycle)
  in $\Ot(n^{2.5})$ time (w.h.p.).
\end{theorem}
Importantly, our algorithm breaks through the best-known strongly polynomial
bound of $O(nm)$ by a polynomial factor
for sufficiently dense graphs with $m=\Omega(n^{1.5+\eps})$.
This provides a partial affirmative answer to the motivating question 2. in the
case of negative-weight SSSP. Moreover, from Theorem~\ref{t:negsssp} one can conclude that the challenge
in obtaining an $O(n^{3-\eps})$ strongly polynomial time bound
for negative SSSP in dense graphs lies beyond the exponential accuracy requirement.


\subsection{Further related work}
The cost of computing on rational numbers on the word RAM has been also a subject
of study in the symbolic computation community. For example, rational
numbers arise when computing the inverse of an integer matrix. Storjohann~\cite{Storjohann15}
showed that one can exactly compute the inverse of an integer matrix in $\Ot(n^3)$ time
which significantly improves upon the naive bound of $\Ot(n^{\omega+1})$ obtained
by multiplying the $\Ot(n^\omega)$ arithmetic operations bound of computing the matrix inverse and the potential $\Theta(n)$ cost
of performing arithmetic operations on rationals.

More recently, the non-trivial cost of arithmetic operations on large integers
on the word RAM arising when counting shortest paths in graphs has been investigated
in~\cite{ChanWX21, ChanWX22}.

Thorup~\cite{Thorup00} studied the non-negative SSSP problem on undirected graphs whose
edge weights are floating point numbers and gave a linear-time algorithm for that case.
Floating point arithmetic is useful in approximate computation, whereas in this paper
we are interested in exact computation exlclusively.

%% file: preliminaries.tex
\section{Preliminaries}\label{s:prelims}
We work with weighted directed graphs $G=(V,E)$. We denote by $\wei_G(e)\in\mathbb{R}$ the weight of
an edge $uv=e\in E$.
If the graph whose edge we refer to is clear from the context, we may sometimes skip the subscript
and write~$\wei(e)$. For simplicity, we do not allow parallel directed edges between the same endpoints of~$G$,
as those with non-minimum weights can be effectively ignored in shortest paths problems.
As a result, we sometimes write $\wei_G(uv)$ or $\wei(uv)$.

For $u,v\in V$, an $u\to v$ path $P$ in $G$ is formally a sequence of vertices $v_1\ldots v_k\in V$, where $k\geq 1$, $u=v_1$, $v=v_k$,
such that $v_iv_{i+1}\in E$ for all $i=1,\ldots,k-1$.
The hop-length $|P|$ of~$P$ equals $k-1$. The length or weight $\wei(P)$ of $P$ is defined
as $\sum_{i=1}^{k-1}\wei_G(v_iv_{i+1})$. $P$ is a \emph{simple path} if $|V(P)|=|E(P)|+1$.
We sometimes view $P$ as a subgraph of $G$ with vertices $\{v_1,\ldots,v_k\}$ and edges (hops) $\{v_1v_2,\ldots,v_{k-1}v_k\}$.

For any $k\geq 0$,
$\dist_G^k(s,t)$ is the minimum length of an $s\to t$ path in $G$ with at most $k$ hops.
A~\emph{shortest $k$-hop-bounded} $s\to t$ path in $G$ is an $s\to t$ path with length $\dist_G^k(s,t)$ and at most~$k$ hops.
We define the $s,t$-distance $\dist_G(s,t)$ as $\inf_{k\geq 0}\dist_G^k(s,t)$.
If $\dist_G(s,t)$ is finite,
there exists a simple $s\to t$ path of length $\dist_G(s,t)$.
Then, we call any $s\to t$ path of length $\dist_G(s,t)$ a \emph{shortest $s,t$-path}.

If $G$ contains no negative cycles, then $\dist_G(s,t)=\dist^{n-1}_G(s,t)$ for all $s,t\in V$.
Moreover, in such a case there exists a \emph{feasible price function} $p:V\to\mathbb{R}$ such
that the \emph{reduced weight}
\linebreak
$\wei_p(e):=\wei(e)+p(u)-p(v)$ is non-negative for all $uv=e\in E$.
For any path $s\to t=P\subseteq G$, the reduced length $\len_p(P)$ (i.e., length wrt.\ weights $\wei_p$) is non-negative and differs
from the original length $\len(P)$ by the value $p(s)-p(t)$ which does not depend on the shape of~$P$.


We sometimes talk about rooted \emph{out-trees} or \emph{out-branchings} $T$ with all edges directed from a parent to a child.
In such a tree $T$, for two vertices $a,b\in V(T)$ such that $a$ is an ancestor of $b$, a \emph{descending path} $T[a\to b]$ is the unique path from $a$ to $b$ in $T$.

\newcommand{\rat}{\mathbb{Q}}
\paragraph{Rational numbers and the word RAM model.} For a positive integer $W$, let us denote by $\rat[W]$
the set of rational numbers $\frac{p}{q}$ with $|p| < W$ and $0<q < W$.

Let us denote by $B = \Theta(\log n)$ the bit length of machine word and for our convenience assume that the rational numbers constituting the input belong to $\Q[2^{B-1}]$, so that both their denominator and numerator (with a sign) fit within one machine word each. We will call such rational numbers \emph{short rationals}. Let us note that $\Q[W_1] \cdot \Q[W_2] , \Q[W_1] / (\Q[W_2] \setminus \{0\}) \subseteq \Q[W_1 \cdot W_2]$ and ${\Q[W_1] + \Q[W_2]}, \Q[W_1] - \Q[W_2] \subseteq \Q[2 \cdot W_1 \cdot W_2]$ (where $C+D$ is the set of numbers that can be represented as $c+d$, where $c \in C$ and $d \in D$ etc.).

For a positive integer $k$, let $\Qq[k]$ denote the set of \emph{$k$-short rationals} defined as $\Q[2^{kB-1}]$. Both the denominator and the numerator (with a sign) of a $k$-short rational fit in $k$ machine words each. Based on the above relations, we conclude that if an arithmetic operation is performed on a $a$-short rational and a $b$-short rational, then the result will be an $(a + b)$-short rational.

As there exist near-linear-time implementations (in the bit-length) of basic operations on integers and of Euclid's algorithm \cite{schonhage1971schnelle} (the naive implementation runs in quadratic time in the bit-length; see
also~\cite{Moller08} for discussion), when dealing with arithmetic operations on $\Q$, we can always assume that they are stored in the irreducible form.
Hence, we can also test the membership to $\Qq[k]$ efficiently.
We sometimes use the term \emph{$b$-bit rational number} when referring to a rational
number from $\rat[2^b]$ whose numerator and denominator together use (at most) $b$ bits.

%% file: overview.tex
\section{Overview of the non-negative rational SSSP algorithm}\label{s:sssp-overview}
\subsection{An $\Ot(nm^{2/3})$-time algorithm}\label{s:dijkstra-easy}
At a high level, our near-optimal non-negative SSSP algorithm
\iffull(Theorem~\ref{t:sssp}, proved in detail in Section~\ref{s:sssp-details}) \fi 
\ifshort(Theorem~\ref{t:sssp}, proved in detail in the full version) \fi  
for short rational weights 
is an implementation of Dijkstra's algorithm using a rather complex
dynamic data structure for performing comparisons requested by Dijkstra's algorithm.

As a warm-up, and to illustrate some of the ideas used to obtain our main result,
in this section we present an implementation of Dijkstra's algorithm for rational data
using a much simpler data structure that already allows polynomial speed-up over the naive $\Ot(mn)$ bound
that trivially estimates the cost of operations on intermediate rational values to be $\Ot(n)$. Recall that all the values
manipulated and compared in Dijkstra's algorithm correspond
to lengths of some simple paths in~$G$, i.e., require $\Ot(n)$ bits to be represented exactly.

Obtaining an $O(n^{2-\eps})$ time bound for rational data could be impossible even for sparse graphs with $m=O(n)$
if we were required to explicitly output all the distances (each of possibly $\Theta(n)$ bits) from the source.
Consequently, we will to stick to a more succinct output representation: a single-source shortest paths tree $T\subseteq G$ itself.
An SSSP tree can be preprocessed in linear time so that it is possible to report
any distance $\dist_G(s,v)$ in $\Ot(n)$ time. This is captured by the following more general
lemma proved in
\iffull
Section~\ref{s:report-distance}.
\fi
\ifshort
the full version.
\fi
\begin{lemma}\label{l:tree-query}
  Let $T$ be an initially empty tree with short rational edge weights.
  There is a data structure supporting constant-time leaf insertions to $T$ and queries computing
  the weight of a descending path in~$T$ such that the query time for a $k$-hop
  path is $\Ot(k)$.
\end{lemma}

\paragraph{Dijkstra's algorithm recap.} Let us first recall how a variant of Dijkstra's algorithm proceeds for an input digraph $G=(V,E)$ with source $s\in V$.
We first make a simplifying assumption that every vertex $v\in V$ is reachable from $s$ in $G$;
this can be achieved by adding, for each $v$, an auxiliary edge $sv$ of a short rational weight $n\cdot (\max_{e\in E}\{\wei_G(e)\})$
unless $sv\in E$. Clearly, this transformation does not affect
the shortest paths from $s$ to the reachable vertices in the original graph.

Dijkstra's algorithm maintains a growing set of \emph{visited vertices} $S\subseteq V$ for which the shortest path from
$s$ has been correctly determined.
Initially $S=\{s\}$.
The set $S$ is iteratively extended by adding a vertex $v^*\in V\setminus S$ minimizing
the value
\begin{equation}\label{eq:est}
  d(v):=\min_{zv\in E:z\in S}\{\dist_G(s,z)+\wei_G(zv)\}.
\end{equation}
One can prove that at that point $d(v^*)=\dist_G(s,v^*)$ holds. Consequently,
if $z^*$ is the minimizer vertex
for $d(v^*)$ in~\eqref{eq:est}, then there exists a shortest path $s\to v^*=P_{v^*}\subseteq G$
whose last edge is $e_{v^*}:=z^*v^*$ and passing through intermediate vertices in $S$ (before adding $v^*$) exclusively.

The vertices $v\in V\setminus S$ are stored in a priority queue $Q$ with the corresponding keys $d(v)$.
Hence, to extend the set $S$, one extracts (and removes) the top element of $Q$.
It is clear that unless $d(v)=\infty$, the value $d(v)$ corresponds to the length of some
$s\to v$ path in $G$, in fact a concatenation of a shortest path from $s$ and a single edge.

When a new vertex $x$ is moved to $S$ (and the distance $\dist_G(s,x)$ is established), the minima in~\eqref{eq:est}
(and thus the keys in the priority queue)
may need to be updated
for vertices~$v$ with an incoming edge from $x$. 
This is also called the \emph{relaxation} of edges $xv$.
Relaxations may lead to
further cost incurred in the priority queue $Q$.

If $Q$ is realized as a simple binary heap, then the bottleneck of the algorithm
lies in $\Ot(m)$ \emph{crucial comparisons} 
of values of the special
form $\dist_G(s,z)+\wei_G(z,v)$
performed while updating the keys $d(v)$ (edge relaxations) and the priority queue $Q$.
If we ran the algorithm without any change on rational data, its total
time cost \emph{not including crucial comparisons} would remain $\Ot(m)$.
However, if we performed crucial comparisons naively, $\widetilde{\Theta}(m)$ of them could take $\Ot(n)$ time each.

\paragraph{A data structure for crucial comparisons.}
While the algorithm proceeds, the edges $e_v$, $v\in S\setminus\{s\}$, defined as before, constitute an
(incremental) out-tree $T_S\subseteq G$ on vertices $S$ (rooted at~$s$) such that
\ifshort
\linebreak
\fi
$\len(T_S[s\to v])=\dist_G(s,v)$ for $v\in S$.
We store $T_S$ in the data structure of Lemma~\ref{l:tree-query}.

Each crucial comparison can be equivalently seen as comparing
two values of the form \linebreak $\len(T_S[s\to z])+\wei_G(zv)$.
Let $T^*$ denote $T_S$ when the algorithm concludes (i.e., when $S=V$).

Let $h\in [1,n]\cap\mathbb{Z}$ and $\gamma>0$ be a parameters to be set later. We will make use of
the \emph{hitting set} trick~\cite{UY91}. Let $H$ be a random subset of $V$ of size
$\lceil \gamma\cdot(n/h)\log{n}\rceil$ vertices of $V$ and also including $s$ in $H$.
We will also call~$H$ the \emph{hitting set}.
We have the following:
\newcommand{\anc}{\alpha}
\begin{fact}\label{l:hitting_basic}
  With high probability (controlled by the constant $\gamma$), for all $v\in V$, the nearest (strict) ancestor $\anc_v\in H$
  of $v$ in $T^*$ is at most $h$ hops apart from $v$ in $T^*$.
\end{fact}
Since $T_S\subseteq T^*$ at all times, the above lemma also holds if we replace $T^*$ with $T_S$.
How is the hitting set helpful when performing crucial comparisons? Suppose
we want to test whether
\begin{equation*}
  \dist_G(s,y)+\wei_G(yu)=\len(T_S[s\to y])+\wei_G(yu)\stackrel{?}{<}\len(T_S[s\to z])+\wei_G(zv)=\dist_G(s,z)+\wei_G(zv),
\end{equation*}
for some $y,z\in S$.
By expressing the path $T_S[s\to y]$ as $T_S[s\to \anc_y]\cdot T_S[\anc_y\to y]$ (and similarly for~$z$), we could equivalently compare:
\begin{equation}\label{eq:delta}
  \underbrace{\len(T_S[s\to \anc_y])-\len(T_S[s\to \anc_z])}_{\Delta_{\alpha_y,\alpha_z}}\stackrel{?}{<}\len(T_S[\anc_z\to z])+\wei_G(zv)-\len(T_S[\anc_y\to y])-\wei_G(yu).
\end{equation}
Observe that by Fact~\ref{l:hitting_basic}, the right-hand side above is
a sum of at most four (possibly negated) lengths of $\leq h$-hop paths in $G$.
Consequently, the right-hand side is an $h$-short rational number.
At the same time, the left-hand side can still be an $\Omega(n)$-bit rational number.
However, note that the number of distinct left-hand sides involved is
proportional to the number of possible pairs $(\anc_y,\anc_z)$, i.e.,
$O(|H|^2)=\Ot((n/h)^2)$.
Our strategy will be to apply preprocessing for each of the pairs $u,v\in H$
to enable near-optimal $\Ot(h)$-time comparisons of the possible
right-hand sides (given explicitly) with values $\Delta_{u,v}$.
For this, we need the following crucial notion.

\newcommand{\bra}{\mathrm{ra}}

\begin{Definition}
  Let $\alpha\in\rat$. Let $b\geq 0$ be an integer. We define the \emph{best $b$-bit rational approximation} $\bra(\alpha,b)$
  to be a pair of (possibly equal) rational numbers $\frac{p_1}{q_1},\frac{p_2}{q_2}$ with $0<q_1,q_2<2^b$ such that
  $\frac{p_1}{q_1}\leq \alpha\leq \frac{p_2}{q_2}$,
  and there is no rational number $\frac{p'}{q'}$ with $0<q'<2^b$ strictly between $\frac{p_1}{q_1}$ and $\frac{p_2}{q_2}$.
\end{Definition}

Note that given $\bra(\alpha,b)=(p_1/q_1,p_2/q_2)$, one can reduce comparing $\alpha$ to any fraction $\beta=p/q$ such that $0<q<2^b$
to comparing $\beta$ with $p_1/q_1$ and $p_2/q_2$ only. Crucially for our needs, we have:

\begin{lemma}\label{l:bra}
  Let $\alpha$ be a $k$-bit rational number. For any positive integer $b=O(\poly{n})$, one can compute $\bra(\alpha,b)$ in $\Ot(k)$ time.
\end{lemma}
The theory of continued fractions gives an elegant proof of Lemma~\ref{l:bra},
\iffull
see Appendix~\ref{s:bra}.
\fi
\ifshort
see the full version.
\fi

With Lemma~\ref{l:bra} in hand, we compute $\bra\left(\Delta_{u,v},\Ot(h)\right)$ for every pair
$u,v\in H$ once both vertices $u,v$ are present in the visited set $S$.
This takes $\Ot(n)$ time since $\Delta_{u,v}$ can be computed from $T_S$ (see Lemma~\ref{l:tree-query})
in $\Ot(n)$ time. Through all pairs $u,v\in H$, the total time cost of computing
best rational approximations is $\Ot(n\cdot (n/h)^2)=\Ot(n^3/h^2)$. Given that, a crucial
comparison~\eqref{eq:delta} can be evaluated in $\Ot(h)$ time
as the right-hand side can be computed in $\Ot(h)$ time by Lemma~\ref{l:tree-query}.

Since the total number of crucial comparisons made by Dijkstra's algorithm is $\Ot(m)$, the running time is
$\Ot(mh+n^3/h^2)$. This bound is optimized for $h=n/m^{1/3}$ and yields an $\Ot(nm^{2/3})$
bound.

\subsection{A hierarchical approach}\label{s:hierarchical-overview}
One bottleneck of our approach so far lied in the $\Theta(n)$ cost of computing
the best rational approximations of a values $\Delta_{u,v}$, $u,v\in H$ (Lemma~\ref{l:bra}).
The approximation itself has size $\Ot(h)$ so a natural idea is to devise a faster
algorithm computing it. However, with no further assumptions on the input
$\alpha$ to Lemma~\ref{l:bra}, it is not clear whether this is possible.

In our case, the required $\widetilde{\Theta}((n/h)^2)$ possible values $\Delta_{u,v}$, $u,v\in H\setminus\{s\}$, are not completely unrelated and
can be defined inductively.
We have:
\begin{align}\label{eq:rozpis}
  \Delta_{u,v}&=\len(T_S[s\to u])-\len(T_S[s\to v])\nonumber\\
              &=(\len(T_S[s\to \alpha_u])-\len(T_S[s\to \alpha_v]))+(\len(T_S[\alpha_u\to u])-\len(T_S[\alpha_v\to v]))\\
              &=\Delta_{\alpha_u,\alpha_v}+(\len(T_S[\alpha_u\to u])-\len(T_S[\alpha_v\to v])).\nonumber
\end{align}
Consequently, $\Delta_{u,v}$ differs from another value of this form
by an $h$-short rational \emph{offset}.
One could hope that it is possible to compute, for $b=\Ot(h)$, the approximation $\bra(\Delta_{u,v},b)$
based on $\bra(\Delta_{\alpha_u,\alpha_v},b)$ and the offset in $\Ot(h)$ time.
Unfortunately, it is not clear how to do that either.

It turns out, however,
that inductive computation is possible if one can accept losing some accuracy.
\iffull
Formally, in Section~\ref{s:bra-add}, we prove the following.
\fi
\ifshort
Formally, in the full version, we prove the following.
\fi
\begin{lemma}\label{l:bra-add}
  Let $\bra(\alpha,b)$ be given. Suppose $\alpha'=p/q$ satisfies $0<q<2^{b'}$  where $b'<b$.
  Then, $\bra(\alpha+\alpha',b-b')$ can be computed in $\Ot(b+\log(1+|\alpha|+|\alpha'|))$ time\footnote{The $\log(1+|\alpha|+|\alpha'|)$ term 
  comes from the fact that the numerator of $\alpha+\alpha'$ may be $|\alpha+\alpha'|$ times larger than the denominator.}.
\end{lemma}
Let us now sketch how Lemma~\ref{l:bra-add} can be taken advantage of. 
Let $h=2^k$ for some positive~$k$.
The accuracy loss in Lemma~\ref{l:bra-add} forces us to use a \emph{hierarchy} of hitting sets $H:=H_k,\ldots,H_\ell$,
where $\ell=\lceil\log_2{n}\rceil$.
For $i<\ell$, let $H_i$ be a random subset of $\Theta((n/2^i)\log{n})$ vertices of $V$
and we put $H_\ell=\{s\}$. As before, w.h.p., $H_i$ hits all the $\min(n,2^i)$-hop descending paths in $T^*$.
Hence,~for any vertex $v\in V$, its nearest ancestor from $H_i$ in $T^*$, denoted $\alpha_{v,i}$,
is at most $2^i$ hops apart from $v$.

Let us set our goal at computing, for all $u,v\in H_i$, the approximation $\bra(\Delta_{u,v},B\cdot 2^{i+2})$. We can
do that inductively for decreasing indices $i<\ell$ (once $u,v$ appear in the set $S$).
By arguing analogously
as in~\eqref{eq:rozpis}, we can conclude that
$\Delta_{u,v}$ differs from some $\Delta_{x,y}$, where $x,y\in H_{i+1}$ by an offset $\beta$
corresponding to a difference of two lengths of $\leq 2^{i+1}$-hop paths, that is, a $2^{i+2}$-short rational
number with denominator bounded by $2^{2^{i+2}\cdot B}$.
By Lemma~\ref{l:bra-add} applied to $\bra(\Delta_{x,y},B\cdot 2^{i+3})$, in $\Ot(B\cdot 2^{i+2})=\Ot(2^i)$ time one can compute
$\bra(\Delta_{u,v},B\cdot 2^{i+2})=\bra(\Delta_{x,y}+\beta,(B\cdot 2^{i+3})-(B\cdot 2^{i+2}))$.
This way, the computation of all approximations $\bra(\Delta_{u,v},\Ot(h))$ for $u,v\in H$, i.e., the bottleneck
in Section~\ref{s:dijkstra-easy}, can be performed in
\begin{equation*}
  \Ot\left(\sum_{i=k}^{\ell}|H_i|^2\cdot 2^i\right)=\Ot\left(\sum_{i=k}^{\ell}\frac{n^2}{2^i}\right)=\Ot(n^2/h)
\end{equation*}
time. By arguing similarly as in Section~\ref{s:dijkstra-easy}, we obtain that Dijkstra's
algorithm for short rational weights can be implemented in improved $\Ot(mh+n^2/h)$ time.
For $h=n/\sqrt{m}$, the bound becomes $\Ot(n\sqrt{m})$ which is already \emph{near-optimal}
for dense graphs, but also far from linear for sparse graphs.
\iffull
\fi

\subsection{The algorithm for sparse graphs}
The bottleneck of our approaches so far lied in constructing a certain
representation of the difference of distances $\dist_G(s,u)-\dist_G(s,v)$ for \emph{all}
pairs $u,v\in H$, where $H$ was a carefully picked subset of $V$.
This approach is inherently super-linear if $|H|>\sqrt{n}$.
On the other hand, using a small subset $H$ led to a large $\Ot(h)$ unit cost
of crucial comparisons. 
To obtain Theorem~\ref{t:sssp}, we eliminate the need for storing all-pairs differences
by exploiting the following two general ideas:
\begin{itemize}
  \item lazy computation of rational approximations of the differences, and
  \item inference of comparison results from the comparisons made so far.
\end{itemize}
Actually, we develop the following data structure maintaining an incremental tree that allows
performing all the crucial comparisons in Dijkstra's algorithm in an entirely black-box way.

\begin{restatable}{lemma}{ldistdifferencequery}\label{l:dist-difference-query}
  Let $c \in \mathbb{N}_{\ge 1}$ be a constant.
  Let $T$ be a~rooted out-tree, initially containing only the root~$s$, with edge weights in $\rat^{(c)}$.
  There exists a Monte Carlo randomized data structure supporting:
  \begin{itemize}
    \item leaf insertions to $T$ (with a specified parent and edge weight); and
    \item answering queries of the form: given two vertices $u, v \in V(T)$ and a~rational number $\frac{p}{q} \in \rat^{(c)}$, compare $\delta_T(s, u) - \delta_T(s, v)$ to $\frac{p}{q}$ and return \emph{smaller}, \emph{equal} or \emph{larger}.
  \end{itemize}
  The data structure processes $n$ leaf insertions and $q$ queries in $\Ot(n + q)$ time. The answers produced are correct with high probability.
\end{restatable}
The data structure works against an adaptive adversary since the correct query results are determined uniquely by the input updates and thus the queries do not reveal any random choices of the data structure (unless a mistake is made, which happens with low probability anyway).

\newcommand{\D}{\mathbb{D}}

To obtain Lemma~\ref{l:dist-difference-query}, the starting point is to reuse the hierarchical approach
of Section~\ref{s:hierarchical-overview}. 
We use a chain of $O(\log{n}/\log\log{n})$ hitting sets of $T$:
\begin{equation*}
  V=L_0\supseteq L_1\supseteq \ldots\supseteq L_t=\{s\},
\end{equation*}
where this time each $L_i$ is $\Theta(\log{n})$ times smaller than the previous set $L_{i-1}$.
The data structure actually consists of a hierarchy of $t+1$ data structures $\D_0,\ldots,\D_t$.
The purpose of $\D_i$ is to efficiently perform comparisons of
a difference of values $\dist_T(s,u)-\dist_T(s,v)$, $u,v\in L_i$, with an $\Ot(n/|L_i|)$-short
rational number $\beta$ given as a query parameter.
Note that the implementation of $\D_t$ boils down to checking the sign of $\beta$,
whereas $\D_0$ is what we actually want to accomplish in Lemma~\ref{l:dist-difference-query}.

The main claim that we prove
\iffull
(Lemma~\ref{lem:dist-inductive-data-structures})
\fi
is that
the time required by $\D_i$ for answering~$q$ queries is $\Ot(q\cdot n/|L_i|)$
plus the cost of performing \emph{only $\Ot(|L_i|)$ queries} (that is, a number independent of $q$) to the data structure $\D_{i+1}$.
Note that in Section~\ref{s:hierarchical-overview}, each comparison of $\dist_G(s,u)-\dist_G(s,v)$ for $u,v\in H_i$ could be thought of being reduced
to a comparison from the previous layer. The crux of our main claim is 
that for most queries to $\D_i$, the answers can be inferred
from the previous answers!
Moreover, the data structures $\D_i$ use separate pools
of random bits and provide unique answers w.h.p., 
and this is why the input to $\D_{i+1}$, even though
generated in a randomized way, can be seen from the point of
view of $\D_{i+1}$ as generated by an oblivious adversary.
At the end, Dijkstra's algorithm performs $m$ queries to $\D_0$,
so the total time used by the data structure $\D_0$
is
\begin{equation*}
  \Ot(m)+\Ot\left(\sum_{i=1}^t n\cdot \frac{|L_{i-1}|}{|L_i|}\right)=\Ot(n+m),
\end{equation*}
as desired.

We now sketch the main ideas behind this approach, focusing only on the implementation of $\D_0$.
Suppose we construct an~incremental out-tree $T$ rooted at $s$.
For every vertex $v \in V$, we maintain a~$\Theta(\log^2 n)$-bit approximation $\delta'_T(s, v)$ of $\delta_T(s, v)$, that is, a~value $\dist'_T(s,v) \in \rat\left[2^{\Theta(\log^2 n)}\right]$ such that
\ifshort
\linebreak
\fi
$|\dist'_T(s, v) - \dist_T(s, v)| < 2^{-\Theta(\log^2 n)}$.
Then, for every $u, v \in V$, one of the following cases holds:
\begin{itemize}
  \item All comparisons between $\dist_T(s, u)$ and $\dist_T(s, v)$ are \emph{easy}.
  That is, for every $\frac{p}{q} \in \rat^{(c)}$, we have $\left|\dist_T(s, u) - \dist_T(s, v) - \frac{p}{q}\right| > 2^{-\Theta(\log^2 n)}$.
  Then, for each comparison as above, we can use the approximations of the distances in lieu of the original distances: in other words, $\dist_T(s, u) - \dist_T(s, v) > \frac{p}{q}$ is equivalent to $\dist'_T(s, u) - \dist'_T(s, v) > \frac{p}{q}$.
  The latter can obviously be evaluated efficiently.
  \item There is a~\emph{difficult} comparison between $\dist_T(s, u)$ and $\dist_T(s, v)$: there exists $\frac{p}{q} \in \rat^{(c)}$ such that
    \begin{equation*}
      |\dist_T(s, u) - \dist_T(s, v) - \frac{p}{q}| \le 2^{-\Theta(\log^2 n)}.
    \end{equation*}
    Then it turns out that there is a unique rational number $\frac{p}{q}\in \rat^{(c)}$ with this property and it can be determined efficiently given $\dist'_T(s, u)$ and $\dist'_T(s, v)$.
    For all other rational numbers $\frac{p'}{q'} \in \rat^{(c)} \setminus \{\frac{p}{q}\}$, $\dist_T(s, u) - \dist_T(s, v) \stackrel{?}{>} \frac{p'}{q'}$ can be tested efficiently based on $\dist'_T(s,u)$ and $\dist'_T(s,v)$.
  Let $u \sim v$ denote that there exists a~difficult comparison between $\dist_T(s, u)$ and $\dist_T(s, v)$, and $u \preceq v$ denote that $\dist'_T(s, u) - \dist'_T(s, v) \leq \frac{p}{q}$.
    \ifshort
    If $u\sim v$, we also say that $u$ and $v$ are \emph{similar}.
    \fi
  Note that $u \preceq v$ implies $u \sim v$.
\end{itemize}

In our algorithm, an~explicit evaluation of $u \stackrel{?}{\preceq} v$ is expensive: it ultimately requires evaluating a~query in $\D_1$.
Therefore, we wish to evaluate $\preceq$ explicitly as few times as possible, and infer the results of the remaining difficult comparisons from the comparisons made so far.

The heart of the algorithm is the following transitivity property for difficult comparisons: suppose that we have three vertices $u, v, w \in V$ such that $u \preceq v$, $v \preceq w$, and $u \sim w$.
Then $u \preceq w$.
\iffull
This property is formalized through the notions of \emph{similarity} and \emph{smallness}, defined in \cref{ssec:dijkstra-real-similarity}.
\fi

Let $H$ now be the graph on $V(T)$ given by the similarity relation $\sim$.
If $H$ is a~clique on $V(T)$ (i.e., all vertices are pairwise similar), then $\preceq$ induces a~total order on $V(T)$; this order can be computed via any comparison-based sorting in $O(n \log n)$ time, plus only $O(n \log n)$ evaluations of~$\preceq$.
Then, all other evaluations of $\preceq$ can be easily inferred from the precomputed order.

By loosening the definition of similarity and smallness slightly, we can generalize this idea to the case where $H$ can be covered using a~small number of vertex-induced subgraphs, so that:
\begin{enumerate}[(i)]
  \item every edge of $H$ is covered by some subgraph,
  \item each subgraph has \emph{weak diameter} $O(\log n)$ (i.e., every pair of vertices in the subgraph is at distance $O(\log n)$ in $H$), and
  \item each vertex belongs to $O(\log n)$ subgraphs.
\end{enumerate}
Such a cover can indeed be constructed efficiently for an~arbitrary graph $H$, see e.g.\ \cite{AwerbuchBCP98}.

Unfortunately, in our setting, the graph $H$ is incremental: we do not know the edges of $H$ beforehand, and we can only find out that $u \sim v$ when a~difficult comparison involving $\delta_T(s, u)$ and $\delta_T(s, v)$ is requested.
In other words, edges are introduced to $H$ throughout the life of $\D_0$.
As a result, in this setting, we must maintain an~\emph{incremental sparse edge cover}: the covering satisfying the properties (i)--(iii) dynamically, with an~additional critical restriction that the total number of changes to the covering
(i.e., the number of insertions/deletions to the vertex sets of individual subgraphs)
throughout the life of $\D_0$ is bounded by $\Ot(n)$.
\iffull
In \cref{l:incremental-nbd-cover}, we propose
\fi
\ifshort
We propose
\fi
a~data structure solving this problem.
The implementation adapts a~clustering technique named \emph{Exponential Start Time Clustering}, proposed by Miller, Peng, Vladu and Xu~\cite{MillerPX13,MillerPVX15}.
Given the incremental sparse edge cover data structure, we can ensure that $\preceq$ is only evaluated $\Ot(n)$ times in total, implying that $\D_1$ is only queried $\Ot(n)$ times even though the number of queries to $\D_0$ may be superlinear.

The above approach, applied (with appropriate parameters) to each data structure $\D_0, \dots, \D_t$ ensures the complexity bounds of the data structure of Lemma~\ref{l:dist-difference-query}.

\section{Overview of the negative rational SSSP algorithm}
In this section we give an overview of our $\Ot(n^{2.5})$-time
algorithm (Theorem~\ref{t:negsssp})
for the negatively-weighted SSSP problem in short rational-weighted graphs.
The algorithm can also detect a negative cycle if one exists, but
for the sake of this overview let us assume there are no negative cycles in~$G$.

Recall that we have developed a very powerful and efficient data structure for performing
comparisons requested by Dijkstra's algorithm (Lemma~\ref{l:dist-difference-query}).
That data structure does not assume anything about the weights it operates
on; in particular these weights can be negative. Hence, we would like to benefit from it as much as possible in the algorithm for negative weights.

\paragraph{A closer look at Dijkstra's algorithm.} To this end, we first identify the possibly minimal tweak one
has apply to Dijkstra's algorithm to make it work for negative weights.
The well-known so-called Johnson's trick~\cite{Johnson} says that, if
one is given a feasible price function $p$ of~$G$,
then Dijkstra's algorithm run on the graph~$G'$ obtained from $G$ by reducing the weights
by $p$ (which makes the weights in $G'$ non-negative) has the same shortest paths as $G$.
The non-negativity of weights is required in Dijkstra's
algorithm only for arguing that the visiting order
of vertices is correct.
This means that as long as the visiting order
when running Dijkstra's algorithm on $G$ is the same as if the algorithm
was run on $G'$, the produced result will be correct.
The latter order is obtained by choosing
a vertex $v\in V\setminus S$ minimizing
\begin{equation*}
  \min_{y\in S}\{\dist_{G'}(s,y)+\wei_p(yv)\}=(\dist_G(s,y)+p(s)-p(y))+\wei(yv)+p(y)-p(v),
\end{equation*}
or equivalently, a vertex $v$ minimizing
\begin{equation*}
  \min_{y\in S}\{\dist_{G}(s,y)+\wei(yv)\}-p(v)=d(v)-p(v).
\end{equation*}
We conclude that Dijkstra's algorithm run on a negatively-weighted $G$ will
produce correct distances if only we shift the standard
key $d(v)$ of $v$ in the priority queue $Q$ by $-p(v)$.

\paragraph{Limited-hop paths.} Unfortunately, computing a feasible price function is not known to be any easier than solving the negatively-weighted SSSP problem itself. However, as we show in
\iffull
Section~\ref{s:scaling},
\fi
\ifshort
the full version,
\fi
the known scaling negatively-weighted SSSP algorithms~\cite{BernsteinNW22, abs-2304-05279} can be applied (in a black-box way) to compute a weaker price function, namely an $\eps$-feasible price function, a notion dating back to~\cite{GabowT89, GoldbergT90}.

\begin{Definition}\label{d:epsfeasible}
Let $G=(V,E)$ be a digraph with edge weights given by $\wei:E\to\rat$. Let $\eps\in\rat_+$. A function $p:V\to\rat$
is called $\eps$-feasible if
for all $uv=e\in E$,
    $\wei_p(e):=\wei(e)+p(u)-p(v)\geq -\eps$.
\end{Definition}
More specifically, an $\eps$-feasible price function with values in $\rat[\Ot(1/\eps)]$ for short rational-weighted
graph can be computed in $\Ot(m\log(1/\eps))$ time on the word RAM.
We set $\eps=2^{-\widetilde{\Theta}(\sqrt{n})}$ in our case.
Hence, computing such an $\eps$-feasible price function $p$ with $\Ot(\sqrt{n})$-bit values takes $\Ot(m\sqrt{n})$ time.

Our previous discussion on a modified processing order in  Dijkstra's algorithm for negative weights is not quite sound in general if $p$ is only $2^{-\widetilde{\Theta}(\sqrt{n})}$-feasible.
However, it turns out that the approach remains correct for short rational-weighted graphs if the shortest paths from $s$ use at most $\sqrt{n}$ hops.
In fact, we require and prove an even stronger property: if we make Dijkstra's algorithm refrain
from processing vertices of $S$ whose distances are not $\sqrt{n}$-short rational numbers, then the computed distances are nevertheless correct for
all the vertices $v$ for which some shortest path from the source has at most $\sqrt{n}$ hops.
More formally, given an arbitrary source $z\in V$, such a variant, that we call the \emph{cut Dijkstra}, computes values $\tilde{d}(v)$ such that if $v$ belongs to the set $C_z$ of vertices
satisfying $\dist_G(z,v)=\dist_G^{\sqrt{n}}(z,v)$, then $\tilde{d}(v)=\dist_G(z,v)$, and otherwise $\tilde{d}(v)$ is the length of \emph{some $\leq\sqrt{n}$-hop path} $z\to v$ (i.e., a distance upper bound). Perhaps surprisingly, from the output of the algorithm, we cannot even tell which vertices belong to $C_z$; we do not need that anyway.

To make use of the cut Dijkstra, we again apply the hitting set trick~\cite{UY91}.
Let $M$ be a random subset of $V$ of size $\Theta(\sqrt{n}\log{n})$. Then, with
high probability, for all pairs $u,v\in V$, $M$ hits some shortest
$u\to v$ path in $G$, unless all such shortest $u\to v$ paths have less than $\sqrt{n}$ hops.
In order to compute distances from $s$ in $G$, we first extend $M$ with the vertex $s$. We then run the cut Dijkstra
from each vertex $z\in M$, obtaining distance estimates $\tilde{d}_z(\cdot)$. Next, we build an auxiliary complete graph $H$
on $M$ so that the weight of an edge $xy$ in $H$ corresponds
to the estimate $\tilde{d}_x(y)$ produced by the cut Dijkstra run from $x$.
Hence, we have
$\wei_H(xy)=\dist_G(x,y)$ if $\dist_G(x,y)=\dist^{\sqrt{n}}_G(x,y)$
and $\wei_H(xy)\geq \dist_G(x,y)$ otherwise.
In a standard way, one shows that for all $y\in M$,
$\dist_H(s,y)=\dist_G(s,y)$ with high probability.
Note that the edge weights in $H$ are $\sqrt{n}$-short rational numbers.
Therefore, single-source shortest paths from $s$ in $H$ can be computed naively
by running the Bellman-Ford algorithm with rational arithmetic cost $\Ot(n)$
in $\Ot(|V(H)|^3\cdot n)=\Ot(n^{2.5})$ time.
Finally, the distances from $s$ in $G$ are computed based
on the equality $\dist_G(s,v)=\min_{z\in M}\{\dist_G(s,z)+\tilde{d}_z(v)\}$
that holds w.h.p. and can be evaluated for each $v$ in $\Ot(n^{3/2})$ time.
Through all $v$, this is again $\Ot(n^{2.5})$ time.

\paragraph{Dealing with costly heap operations in the cut Dijkstra.}
We have not yet analyzed the running time of the cut Dijkstra given
an $\eps$-feasible price function $p$.
Roughly speaking, the time cost of the cut Dijkstra excluding
the time needed for selecting the vertex minimizing $d(v)-p(v)$
is no more than the running time of the algorithm of Theorem~\ref{t:sssp} for non-negative data, i.e., $\Ot(m)$.
Unfortunately, efficiently selecting the smallest key $d(v)-p(v)$
over all $v\in V\setminus S$ is highly non-trivial.
The challenge lies in the fact that the values $p(v)$
are $O(\sqrt{n})$-short rationals and thus Lemma~\ref{l:dist-difference-query} cannot be applied. If the priority
queue was implemented naively (e.g., using a heap),
then the $\Ot(m)$ comparisons in the queue could take $\Omega(m\sqrt{n})$ time.
But recall that we need to run the cut Dijkstra $|M|=\Ot(\sqrt{n})$ times,
so this would lead to a potentially cubic $\widetilde{\Theta}(mn)$ total time.

To circumvent this problem, we carefully analyze what happens when
a processed vertex $v\in S$ relaxes its outgoing edges $vu_1,vu_2,\ldots,vu_l$,
where $u_1,\ldots,u_l\in V\setminus S$.
We argue that if we sort the values $\wei(vu_i)-p(u_i)$ so that $\wei(vu_1)-p(u_1)\leq \ldots \leq\wei(vu_l)-p(u_l)$,
then the key of $u_i$ in the priority queue does not need to be updated until $i-1$ more vertices
of $S$ are processed (unless $d(u_i)$ is updated in the meantime).
Our implementation of the cut Dijkstra delays key updates using this observation.
We subsequently prove that the maximum number of such delayed key updates performed in a single run of the cut Dijkstra is bounded by $O(n^{3/2})$.
This is achieved
by analyzing an interesting game
\iffull
(Lemma~\ref{lem:game}).
\fi
\ifshort
(see the full version).
\fi
This way, we reduce the total cost of priority queue key updates to $\Ot(n^2)$.
Note that this is significantly less than $\Ot(m\sqrt{n})$ for dense graphs.

It remains to show how the values $\wei(vu_i)-p(u_i)$ can be efficiently sorted
when processing the vertex $v\in S$.
A reasonable sorting algorithm will perform $\Ot(l)=\Ot(n)$ comparisons
of the form $\wei(vu_i)-\wei(vu_j)<p(u_j)-p(u_i)$ in that case.
Note that the left hand side in this comparison is an $O(1)$-short rational,
whereas the right-hand side is an $\Ot(\sqrt{n})$-short rational.
So a naive implementation would spent $\Ot(\sqrt{n})$ time per comparison,
which would again lead to prohibitive $\Ot(m\sqrt{n})$ cost of the cut Dijkstra.
However, we can again apply one of our early ideas here!
Namely, given best $\Ot(1)$-bit rational approximations of the $n^2$
values $p(y)-p(x)$ for all $x,y\in V$, the total time spent in sorting
per cut Dijkstra run would be only $\Ot(m)$.
Since the values $p(\cdot)$ are $\Ot(\sqrt{n})$-bit, such an additional
preprocessing step takes $\Ot(n^{2.5})$ time by Lemma~\ref{l:bra}.
The stored approximations are shared among multiple cut Dijkstra runs,
and enable running a single cut Dijkstra in $\Ot(n^2)$ time.
Since the number of runs is $|M|=\Ot(\sqrt{n})$, the total cost
is once again bounded by $\Ot(n^{2.5})$.

%% file: fast-sssp.tex
\newcommand{\tudu}{\textcolor{red}{[TODO]}\xspace}

In this section, we will prove \cref{t:sssp}; that is, show a~randomized algorithm solving the single-source shortest paths problem in the regime of non-negative edge weights.
For convenience, we restate it here, using the notation introduced in the Preliminaries.

\begin{theorem}[Restatement of \cref{t:sssp}]
  \label{thm:rational-dijkstra}
  Fix a~constant $c \in \mathbb{N}_{\ge 1}$.
  Suppose we are given a~directed weighted graph $G$ with non-negative edge weights from $\rat^{(c)}$, and a~source $s \in V(G)$.
  Suppose every vertex of $G$ is reachable from $s$.

  Then, with high probability, in time $\Ot(n + m)$ we can compute a~single-source shortest paths tree from $s$, i.e., an~out-tree $T$ rooted at $s$ such that $V(T) = V(G)$, $E(T) \subseteq E(G)$ and for every $v \in V(G)$, we have that $\delta_T(s, v) = \delta_G(s, v)$.
\end{theorem}

The main technical tool of this section is a data structure maintaining an~incremental, edge-weighted out-tree $T$ rooted at $s$, supporting the comparisons between the distances from $s$ to the vertices of $T$:

\ldistdifferencequery*

We stress that both the edge weights and the rational numbers in the queries might be negative; note that the distances $\delta_T(s, \cdot)$ are still defined correctly: $s$ is the root of the out-tree $T$, so for every vertex $v \in V(T)$, there exists a~unique path in $T$ from $s$ to $v$.

At this point of time, we present an~efficient algorithm for non-negative single-source shortest paths, assuming \cref{l:dist-difference-query}.

\begin{proof}[Proof of \cref{thm:rational-dijkstra}]
  We adapt Dijkstra's algorithm to the setting of rational numbers.
  Recall that throughout the algorithm, we maintain a~growing set of visited vertices $S \subseteq V$, where initially $S = \{s\}$.
  Then, we iteratively select a~vertex $v^\star \in V \setminus S$ with the minimum value of
  \[ d(v):=\min_{zv\in E:z\in S}\{\dist_G(s,z)+\wei_G(zv)\} \]
  and add $v^\star$ to $S$.
  This is implemented by maintaining a~priority queue $Q$ supporting: (1) adding the elements of the form $\dist_G(s,z)+\wei_G(zv)$ for $z \in S$, $zv\in E$, and (2) extracting the smallest element of $Q$.
  
  At the start of the algorithm, we initialize the data structure of \cref{l:dist-difference-query} with a~rooted out-tree, initially only containing the root $s$, and supporting the edge weights and queries in $\rat^{(2c)}$; the data structure maintains the shortest paths tree $T$ on the vertices from $S$.
  Whenever a~new vertex~$v$ is introduced to $S$ through an~edge $(u, v)$ of weight $w$, we add that edge to the tree $T$.
  Thus, a~vertex is added as a~leaf to $T$ at most $n$ times.

  Then, the priority queue $Q$ is implemented as follows: whenever an~element of the form ${\dist_G(s,z)+\wei_G(zv)}$ is to be added to the queue, we emplace a~pair $(z, \wei_G(zv))$.
  In order to compare two pairs $(z_1, w_1)$ and $(z_2, w_2)$ in the queue ($z_1, z_2 \in S$, $w_1, w_2 \in \rat^{(c)}$), we query the data structure of \cref{l:dist-difference-query} to compare $\delta_T(s, z_1) - \delta_T(s, z_2)$ with $w_2 - w_1 \in \rat^{(2c)}$.
  With high probability, the data structure returns the correct answer, so we correctly decide whether $\delta_T(s, z_1) - \delta_T(s, z_2) < w_2 - w_1$; or equivalently, whether $\delta_T(s, z_1) + w_1 < \delta_T(s, z_2) + w_2$.
  
  Assuming that $Q$ is implemented using a~binary heap, the priority queue performs $\Ot(m)$ crucial comparisons; thus, the data structure of \cref{l:dist-difference-query} is queried $\Ot(m)$ times.
  Therefore, the data structure processes all operations correctly with high probability and in $\Ot(n + m)$ time.
  Hence, our algorithm also outputs the correct result with high probability and in near-linear time.
\end{proof}

The remaining part of this section is devoted to the proof of \cref{l:dist-difference-query}.
The structure of this section is as follows.
In \cref{ssec:incremental-sparse-edge-covers}, we introduce a~data structure for \emph{incremental sparse edge covers}, maintaining a well-structured covering of the set of vertices of an~incremental graph $H$.
Following that, in \cref{ssec:dijkstra-real-similarity}, we define the notion of \emph{similarity} of real numbers: intuitively, two real numbers $x$, $y$ are similar if their difference $x - y$ is very close to a~short rational number.
Then, we prove the properties of similarity that will be exploited in the design of the data structure from \cref{l:dist-difference-query}.
Finally, in \cref{ssec:dijkstra-distance-cmp}, we use the introduced toolchain to produce an~efficient data structure for distance comparisons, proving \cref{l:dist-difference-query}.

\subsection{Incremental sparse edge covers}
\label{ssec:incremental-sparse-edge-covers}

The first component we introduce is a~data structure maintaining, for an~incremental undirected graph, a~dynamic family of subsets of vertices which we call a \emph{sparse edge cover}:


\begin{lemma}\label{l:incremental-nbd-cover}
  Let $H$ be an~initially empty $n$-vertex simple undirected graph, updated by edge additions, and let $\lambda > 0$.
  There is a~non-adaptive\footnote{Non-adaptive, that is, we assume that the sequence of queries to the data structure is predetermined, but unknown to the implementation of the data structure.} randomized data structure maintaining a~covering $V(H) = A_1 \cup A_2 \cup \dots \cup A_k$, updating the covering on each edge insertion by: adding or removing empty sets, and adding or removing vertices from the sets.
  On each edge insertion, the data structure returns the sequence of performed updates.

  With high probability (controlled by the constant $\lambda$), after each graph update, the covering satisfies the following properties:
  \begin{itemize}
    \item \emph{(edge covering condition)} for all $uv \in E(H)$, there exists $i \in \{1, \dots, k\}$ such that $u, v \in A_i$;
    \item \emph{(weak diameter condition)} for all $i \in \{1, \dots, k\}$ and $u, v \in A_i$, we have $\delta_H(u, v) \leq \lambda \log n$;
    \item \emph{(sparsity condition)} each $v \in E(H)$ is in at most $\lambda \log n$ sets $A_i$.
  \end{itemize}
  With high probability (controlled by $\lambda$), the data structure processes $m$ edge additions in $\Ot(n+m)$ time and issues a total of $\Ot(n)$ single-element updates to the individual sets of the covering.
\end{lemma}

We stress that the covering is updated $\Ot(n)$ times, regardless of the value of $m$.
In particular, when $m \gg n$ edges are inserted to the graph, the covering remains unchanged for most of the edge insertions.

We prove \cref{l:incremental-nbd-cover} in the remaining part of the section.

\paragraph*{Exponential Start Time Clustering.}
The implementation of the data structure crucially relies on the method of Exponential Start Time Clustering by Miller et al.~\cite{MillerPX13,MillerPVX15}.
Here, we introduce the discrete variant of the method and state its various properties.
The continuous versions of these properties are proved in \cite{MillerPX13,MillerPVX15}, so we provide the proofs of the statements in \cref{s:proofs-estc}.

\newcommand{\Geo}{\mathrm{Geo}}
For a~real number $p \in (0, 1)$, the geometric distribution $\Geo(p)$ is the integer-valued probability distribution of the number of failures in independent Bernoulli trials with success probability $p$ before the first success; that is, for any random variable $X \sim \Geo(p)$ and $k \in \mathbb{N}_{\ge 0}$, we have $\Pr[X \geq k] = (1 - p)^k$.
For our convenience, for a~real $\alpha > 0$, we will consider the distribution $\Geo(1 - e^{-\alpha})$. That is, for a~random variable $X \sim \Geo(1 - e^{-\alpha})$, we have $\Pr[X \geq k] = e^{-\alpha k}$.

Consider the following clustering algorithm partitioning a~graph $H$ into vertex-disjoint clusters $C_1, \dots, C_t$:

\begin{algorithm}[H]
	\hspace*{\algorithmicindent} \textbf{Input:} unweighted undirected graph $H = (V, E)$, parameter $\alpha > 0$. \\
	\hspace*{\algorithmicindent} \textbf{Output:} vertex partitioning $V = C_1 \cup \dots \cup C_t$. 
	\begin{algorithmic}[1]
			
		\caption{Discrete Exponential Start Time Clustering} \label{alg:clustering}
		\State For each $v \in V$, draw $b_v \sim \Geo(1 - e^{-\alpha})$ independently at random.
		\State For each $v \in V$, assign $v$ to the cluster centered at $u$, where $u = \arg \min_{w \in V} ( \dist_H(v, w) - b_w )$.
				If multiple values of $w$ minimize the value $\dist_H(v, w) - b_w$, pick one arbitrarily.
		\State Return the set of clusters created this way.
	\end{algorithmic}
\end{algorithm}

Note that we explicitly do not introduce any tie-breaking rules in Step 2 above (even though due to this decision, some of the produced clusters could become disconnected); this choice will be essential to the design of the incremental sparse edge cover data structure.
Also remark that the choice of the values $b_v$ is not dependent on the graph given as input.

We follow with the properties of the partitioning $C_1 \cup \dots \cup C_k$ produced by the Discrete Exponential Start Time Clustering.
The proofs are deferred to \cref{s:proofs-estc}.

\begin{lemma}
  \label{lem:clustering-diameter}
  Let $k \in \mathbb{N}_{\ge 1}$.
  Then, with probability at least $1 - \frac{1}{n^{k-1}}$, we have $b_v \leq \frac{k \log n}{\alpha}$ for all $v \in V$.
  Moreover, if $b_v \leq \frac{k \log n}{\alpha}$ for all $v \in V$, then independently of the tie-breaking rule
  , for every $i \in \{1, \dots, t\}$ and every pair of elements $u, v \in C_i$, we have $\dist_H(u, v) \leq \frac{k \log n}{\alpha}$.
\end{lemma}

\begin{lemma}
  \label{lem:edge-probab-clustering}
  Let $uv \in E$ be an~edge of $H$.
  The probability that both endpoints of the edge are in the same cluster is at least $e^{-2\alpha}$, independently of the tie-breaking rule
\end{lemma}

\newcommand{\Hshift}{H^{\mathrm{shift}}}

Finally, let $b_{\max} = \max_{v \in V} b_v$.
Consider the \emph{shifted graph} of $H$, denoted $\Hshift$, constructed as follows: introduce to $H$ a~new vertex $s$, called a~\emph{source}.
Then, for each $v \in V$, connect $s$ with $v$ with a~path $P_v$ of length $b_{\max} + 1 - b_v$. For convenience, let $x_v$ be the neighbor of $s$ on $P_v$.
It is easy to verify that:

\begin{observation}
\label{obs:shift-graphs-and-clustering}
In the Discrete Exponential Start Time Clustering, $v \in V$ can be assigned to the cluster centered at $u \in V$ if and only if there exists some shortest path from $s$ to $v$ with prefix $P_u$; or equivalently, $x_u$ lies on some shortest path from $s$ to $v$.
\end{observation}

\paragraph*{Incremental low-depth BFS trees.}
We will need to maintain a low-depth BFS tree of an unweighted graph under edge insertions. 
More formally, we use the following variant of the classical Even-Shiloach tree~\cite{EvenS81}\footnote{The Even-Shiloach tree is usually stated
in the decremental setting. Its incremental is conceptually even simpler.} suited to our needs.
\begin{lemma}{\textup{\cite{EvenS81}}}
	\label{lem:incremental-second-vertex}
	Let $G$ be an initially connected unweighted dynamic undirected graph, updated by edge insertions only, and whose diameter never exceeds $D$. Let $s \in V(G)$.
	It is possible to maintain, for each vertex $v \neq s$, a~vertex $\beta_v \neq s$ that is a~neighbor of $s$ on some shortest path from $s$ to $v$.
  The initialization and the following sequence of $q$ insertions is processed in total time $O(n+(q+m)D)$, and the values $\beta_v$ are updated at most $O(nD)$ times in total.
\end{lemma}
\begin{proof}
	On initialization, compute any BFS tree $T$ rooted at $s$ and store the resulting values $\beta_v$.
  
  When an edge $uv$ is inserted to $G$, $\dist_G(s,v)$ can only change to $\dist_G(s,u)+1$
  if that value is less than the distance to $v$ before the insertion.
  If this is not the case, the tree does not need to be updated.

  If the distance to $v$ drops, then the distances to its adjacent vertices can also drop,
  and so on.
  Hence, to perform the update, we can mimic breadth-first search. We initialize the queue $Q$ of vertices
  whose distances drop due to the insertion and repeatedly process the vertices $y\in Q$, possibly detecting new distance drops (causing queue pushes)
  by iterating through $y$'s incident edges,
  until the queue $Q$ empties.
  While establishing new distances, we can also easily maintain the values $\beta_v$ based
  on the respective values of the parents in the tree $T$ being rebuilt top-down.
  
  Clearly, throughout the entire sequence of edge insertions, a vertex $u$ is only
  put in the queue (and its neighborhood is explored in $O(\deg(v))$ time) if its distance drops which can
  happen at most $D$ times.
  At the same time, the value $\beta_u$ can only change if $u$ is put on the queue, i.e.,
  $O(D)$ times.
  It follows that the total update time through a sequence of $q$ insertions
  is bounded by $O(q+D\cdot \sum_{v\in V}\deg(v))=O((m+q)D)$.
\end{proof}

\paragraph*{Incremental Exponential Start Time Clustering.}
We now show how the clustering produced by \cref{alg:clustering} works in the incremental setting, where we are given an~$n$-vertex graph $H$ updated by edge insertions.
More formally, we prove the following statement:

\begin{lemma}
	\label{lem:single-cluster}
	Let $H$ be an~initially empty $n$-vertex undirected graph, updated by edge additions. Let also $k \ge 1$.
	We can maintain in a~non-adaptive way a~clustering $V(H) = C_1 \cup \dots \cup C_t$, updating the covering on each edge insertion by: adding or removing empty sets, and adding or removing vertices from the sets.
	On each edge insertion, we return the sequence of performed updates.
	Then:
	\begin{itemize}
	  \item independently for each snapshot of the dynamic graph and each edge $uv$ present in the snapshot, the vertices $u$ and $v$ belong to the same cluster with probability at least $e^{-2}$;
	  \item with probability at least $1 - \frac{1}{n^{k - 1}}$, in all snapshots of the dynamic graph, all clusters $C_i$ and all $u, v \in C_i$, we have $\dist_H(u, v) \leq k \log n$.
	\end{itemize}
	With probability at least $1 - \frac{1}{n^{k-1}}$, the data structure processes $m$ edge additions in time $\Ot(n + m)$ and updates the covering $\Ot(n)$ times.
\end{lemma}

For convenience, fix $\alpha \coloneqq 1$.
On initialization, draw the integers $b_v \sim \Geo(1 - e^{-\alpha})$ for each $v \in V$ independently at random, and put each vertex $v \in V(H)$ into a~one-element cluster centered at $v$.
Let $b_{\max}$ be the maximum value of $b_v$.
By \cref{lem:clustering-diameter}, we have $b_{\max} \leq k \log n$ with probability at least $1 - \frac{1}{n^{k-1}}$.
We construct the shifted graph $\Hshift$ of $H$ in time $\Ot(n)$; note that initially $|V(\Hshift)| = \Ot(n)$ and $|E(\Hshift)| = \Ot(n)$ with high probability. Then initialize a~data structure of \cref{lem:incremental-second-vertex} maintaining $\Hshift$.
The data structure maintains, for each vertex $v \in V(\Hshift) \setminus \{s\}$, a~neighbor $\beta_v$ of $s$ such that $\beta_v$ lies on some shortest path from $s$ to $v$.
Note that by the construction of $\Hshift$, for each $v \in V(H)$ we have that $\beta_v = x_u$ for some $u \in V(H)$; and by \cref{obs:shift-graphs-and-clustering}, $v$ can then be assigned to the cluster centered at $u$.

Now, each edge insertion to $H$ implies an~analogous edge insertion to $\Hshift$; after the insertion, some of the values $\beta_v$ for some vertices $v \in V$ may change; on each modification of $\beta_v$ (say from $x_{u_1}$ to $x_{u_2}$), we move $v$ from the cluster centered at $u_1$ to the cluster centered at $u_2$.

It remains to prove the correctness and efficiency bounds of \cref{lem:single-cluster}.
By \cref{lem:clustering-diameter}, we have $b_{\max} \leq k \log n$ with probability at least $1 - \frac{1}{n^{k-1}}$, so \cref{lem:incremental-second-vertex} applies with $D = O(\log n)$ with high probability and therefore any sequence of $m$ edge insertions, starting from an~empty graph takes time $\Ot(n + m)$; and moreover, the values $\beta_v$ are updated at most $\Ot(n)$ times in total.
Therefore, the total number of elementary changes to the clustering is upper-bounded by $\Ot(n)$.

Note that the choice of the parameters $b_v$ in the Exponential Start Time Clustering does not depend on the input graph.
Since we assume the data structure is non-adaptive, we may assume that the sequence of updates to the graph is predetermined; so after each update, the current clustering of the graph is a~clustering that could be produced by \cref{alg:clustering} when run on each snapshot of the graph from scratch with some tie-breaking strategy.
Therefore:
\begin{itemize}
  \item For some fixed edge $uv$ in some fixed snapshot of $H$, we have by \cref{lem:edge-probab-clustering} that the probability that both $u$ and $v$ are in the same cluster $C_i$ is at least $e^{-2}$;
  \item By \cref{lem:clustering-diameter}, we have $b_{\max} \le k \log n$ with probability at least $1 - \frac{1}{n^{k-1}}$. Provided that this inequality holds, then again by \cref{lem:clustering-diameter}, in every snapshot of the graph and every cluster $C_i$, we have that $\dist_H(u, v) \leq k \log n$ for each pair of vertices $u, v \in C_i$.
\end{itemize}

This concludes the proof of \cref{lem:single-cluster}.

\paragraph*{Multiple independent clusterings.}
We are now ready to prove \cref{l:incremental-nbd-cover}.

\begin{proof}[Proof of \cref{l:incremental-nbd-cover}]
  Suppose that $\lambda > 50$ and set $k = \lambda$.
  Let $T \coloneqq \lambda \log n$ and create $T$ independent instances of Incremental Exponential Start Time Clustering (\cref{lem:single-cluster}); the $i$th instance maintains its own clustering $\mathcal{C}_i = \{C^i_1, C^i_2, \dots, C^i_{t_i}\}$.
  Our data structure maintains the covering $\mathcal{C} \coloneqq \bigcup_{i=1}^T \mathcal{C}_i$.  
  
  On each edge insertion, we relay the insertion to each instance of Incremental Exponential Start Time Clustering and apply each update received from each instance to the covering $\mathcal{C}$.
  
  We verify that:
  \begin{itemize}
    \item The edge covering condition holds in $\mathcal{C}$:
      fix a~snapshot of the incremental graph and an~edge $uv$ of the graph.
      Since the clusterings $\mathcal{C}_i$ are independent, the probability that $u$ and $v$ are in different clusters in each clustering is at most
      \[ (1 - e^{-2})^T \leq (e^{-1/10})^T = e^{(-1/10) k \log n} = n^{-k / 10}. \]
      
      Since the number of snapshots of the graph is bounded by $n^2$ (note that the graph is simple, so we may add at most one edge between any pair of vertices) and the number of edges is bounded by $n^2$ in each instance, we infer from the union bound that the probability that in any snapshot, the endpoints of any edge are assigned to different clusters in each clustering is at most
      \[ n^4 \cdot n^{-k / 10} = \frac{1}{n^{k/10 - 4}}. \]
      Since $k > 50$ and $k$ is chosen arbitrarily, we conclude that the edge covering condition is satisfied with high probability.
      
    \item The weak diameter condition holds in $\mathcal{C}$. By \cref{lem:single-cluster}, this condition holds for all snapshots of the dynamic graph in the $i$th instance of Incremental Exponential Start Time Clustering with probability at least $1 - \frac{1}{n^{k-1}}$.
    Since $T = k \log n$, we get by union bound that the condition holds for all clusterings with probability at least $1 - \frac{k\log n}{n^{k - 1}} > 1 - \frac{1}{n^{k-2}}$ for $n$ large enough.
    
    \item The sparsity condition holds trivially by construction.
    
    \item With high probability (at least $1 - \frac{1}{n^{k-1}}$), the $i$th instance of the data structure from \cref{lem:single-cluster} processes $m$ edge additions in time $\Ot(n + m)$ and updates its clustering $\mathcal{C}_i$ at most $\Ot(n)$ times.
    Again using union bound, we see that with high probability, all instances process $m$ edge additions in total time $\Ot(n + m)$ and issue at most $\Ot(n)$ updates in total to the covering~$\mathcal{C}$.
  \end{itemize}
  Since all required conditions are satisfied with high probability, this concludes the proof.
\end{proof}

\subsection{Similarity of real numbers}
\label{ssec:dijkstra-real-similarity}

We now introduce the notion of similarity of real numbers.
Assume $b, \ell \geq 0$ are integer parameters.
We say that two real numbers $x, y \in \R$ are \emph{$(b, \ell)$-similar} (denoted $x \sim_b^\ell y$) if there exists a~rational number $\frac{p}{q} \in \rat[2^b]$ such that
\[ \left| x - y - \frac{p}{q} \right| \leq 2^{-\ell}. \]
In other words, $x$ and $y$ are $(b, \ell)$-similar if and only if their difference is sufficiently close to some rational number in $\rat[2^b]$.
Naturally, this definition is symmetric: whenever $x$ and $y$ are $(b, \ell)$-similar, then $y$ and $x$ are $(b, \ell)$-similar.
Also, if $x$ and $y$ are $(b, \ell)$-similar, then also $x$ and $y$ are $(b', \ell')$-similar whenever $b' \geq b$ and $\ell' \leq \ell$.

Observe that if $\ell$ is sufficiently large compared to $b$, then the rational number $\frac{p}{q}$ above is defined uniquely:
\begin{lemma}
  \label{l:frac-unique-similarity}
  Assume that $\ell \geq 2b + 2$ and let $x, y \in \R$.
  If there exist two rational numbers $\frac{p_1}{q_1}, \frac{p_2}{q_2} \in \rat[2^b]$ such that $\left| x - y - \frac{p_1}{q_1} \right| \leq 2^{-\ell}$ and $\left| x - y - \frac{p_2}{q_2} \right| \leq 2^{-\ell}$, then $\frac{p_1}{q_1} = \frac{p_2}{q_2}$.
\end{lemma}
\begin{proof}
  Observe that
  \[ \left| \frac{p_1}{q_1} - \frac{p_2}{q_2} \right| \leq \left| x - y - \frac{p_1}{q_1} \right| + \left| x - y - \frac{p_2}{q_2} \right| \le 2^{-(\ell - 1)}. \]
  Since $q_1, q_2 > 0$, we equivalently have that
  \[ |p_1q_2 - p_2q_1| \leq 2^{-(\ell - 1)} \cdot q_1q_2. \]
  The left-hand side of the inequality is nonnegative and integral, while the right-hand side is upper-bounded by $\frac12$ from the facts that $\ell \ge 2b+2$ and $q_1, q_2 < 2^b$.
  Thus $p_1q_2 = p_2q_1$, or equivalently $\frac{p_1}{q_1} = \frac{p_2}{q_2}$.
\end{proof}

Therefore, for $\ell \geq 2b+2$, we introduce the following notation.
Given two real numbers $x, y \in \R$, we say that $x$ is \emph{$(b, \ell)$-smaller} than $y$ (denoted $x \preceq_b^\ell y$) when $x \sim_b^\ell y$ and for the unique rational number $\frac{p}{q} \in \rat[2^b]$ such that $\left| x - y - \frac{p}{q} \right| \le 2^{-\ell}$, we have that $x - y \leq \frac{p}{q}$.
We remark that if $x \preceq_b^\ell y$ and $y \preceq_b^\ell x$, then actually $x - y = \frac{p}{q}$ for the unique $\frac{p}{q} \in \rat[2^b]$ with $\left| x - y - \frac{p}{q} \right| \le 2^{-\ell}$.
In this case, we will say that $x$ and $y$ are \emph{$(b, \ell)$-equal} (denoted $x \approx_b^\ell y$).

\medskip
The notion of similarity enjoys the following \emph{weak} variant of equivalence:
\begin{lemma}
  \label{l:frac-similarity-equivalence}
  Let $b_1, b_2 \geq 0$ and $\ell \geq 1$.
  Suppose for three real numbers $x, y, z \in \R$ that $x \sim_{b_1}^{\ell} y$ and $y \sim_{b_2}^{\ell} z$.
  Then $x \sim_{b_1 + b_2 + 1}^{\ell - 1} z$.
\end{lemma}
\begin{proof}
  Let $\frac{p_1}{q_1} \in \rat[2^{b_1}]$, $\frac{p_2}{q_2} \in \rat[2^{b_2}]$ be such that $\left|x - y - \frac{p_1}{q_1}\right| \le 2^{-\ell}$ and $\left|y - z - \frac{p_2}{q_2}\right| \le 2^{-\ell}$.
  Then the rational number $\frac{p}{q} \coloneqq \frac{p_1}{q_1} + \frac{p_2}{q_2}$ belongs to $\rat[2^{b_1 + b_2 + 1}]$.
\end{proof}
Intuitively, when $x$ and $y$ are similar, and $y$ and $z$ are similar, then also $x$ and $z$ are similar, albeit possibly with worse parameters $b$ and $\ell$.

The most important result in this section is the following weak variant of transitivity for the relation of $(b,\ell)$-smallness: given that $\ell$ is sufficiently large in terms of $b$, from the facts that $x$ is $(b,\ell)$-smaller than $y$, which in turn is $(b,\ell)$-smaller than $z$, and additionally that $x$ and $z$ are $(b,\ell)$-similar, we can infer that $x$ is $(b,\ell)$-smaller than $z$.
Note that in the conclusion of the lemma, the parameters with which $x$ is smaller than $z$ do \emph{not} worsen.
\begin{lemma}
  \label{l:frac-smaller-transitivity}
  Let $b, \ell \ge 0$ be such that $\ell \geq 4b + 5$.
  Suppose for three real numbers $x, y, z \in \R$ that $x \preceq_b^\ell y$, $y \preceq_b^\ell z$ and $x \sim_b^\ell z$.
  Then $x \preceq_b^\ell z$.
\end{lemma}
\begin{proof}
   Let $\frac{p_1}{q_1} \in \rat[2^b]$, $\frac{p_2}{q_2} \in \rat[2^b]$ be such that $\left|x - y - \frac{p_1}{q_1}\right| \le 2^{-\ell}$ and $\left|y - z - \frac{p_2}{q_2}\right| \le 2^{-\ell}$; moreover, $x - y < \frac{p_1}{q_1}$ and $y - z < \frac{p_2}{q_2}$.
   Let also $\frac{p_3}{q_3} \in \rat[2^b]$ be such that $\left| x - z - \frac{p_3}{q_3} \right| \le 2^{-\ell}$.
   Then, letting $\frac{p}{q} \coloneqq \frac{p_1}{q_1} + \frac{p_2}{q_2} \in \rat[2^{2b + 1}]$, we have that
   $\left| x - z - \frac{p}{q} \right| \le 2^{-(\ell - 1)}$ and $x - z < \frac{p}{q}$.

   Applying the fact that $(\ell - 1) \geq 2(2b + 1) + 2$, we get from \cref{l:frac-unique-similarity} that $\frac{p}{q}$ is the unique fraction in $\rat[2^{2b + 1}]$ with the property that $\left| x - z - \frac{p}{q} \right| \le 2^{-(\ell - 1)}$.
   Since $\frac{p_3}{q_3} \in \rat[2^{2b + 1}]$ and $\left| x - z - \frac{p_3}{q_3} \right| \le 2^{-\ell} < 2^{-(\ell - 1)}$, we deduce that $\frac{p_3}{q_3} = \frac{p}{q}$.
   The statement of the lemma follows from observing that $x - z < \frac{p}{q} = \frac{p_3}{q_3}$.
\end{proof}

\subsection{Data structure for distance comparisons}
\label{ssec:dijkstra-distance-cmp}

This section is dedicated to the proof of \cref{l:dist-difference-query}.
Henceforth, fix $k > 0$; our aim is to describe the data structure that processes any sequence of $n$ leaf insertions and $q$ queries correctly in time $\Ot(n + q)$ with probability at least $1 - n^{-k}$.
Let $\lambda > 1$ be the constant ensuring the correctness of the data structure from \cref{l:incremental-nbd-cover} with probability at least $n^{-(k + 1)}$.
Similarly, let $\gamma > 1$ be the constant from \cref{l:hitting_basic} ensuring the following: assume that, given an~$n$-vertex out-tree $T$ and a~real $h \ge 1$, we sample a~random subset $H$ of $\left\lceil n / h \right\rceil$ nodes of $T$ containing the source.
Then, with probability at least $n^{-(k+1)}$, for all $v \in V(T)$, the nearest ancestor of $v$ in $H$ is at most $\gamma \cdot h \log n$ hops away from $v$.
Finally, recall from the statement of the lemma the constant $c > 1$ with the property that all short rational numbers supplied to the data structure are in $\rat^{(c)} = \rat[2^{cB - 1}]$.

Assume that the upper bound $n \gg 1$ (with $n$ large enough) on the number of leaf insertions is known at the time of the initialization of the data structure\footnote{A standard data structure design technique allows us to make this assumption: initialize the data structure with $n = 2$. Whenever the data structure undergoes the $(n+1)$th insertion, reinitialize the entire data structure with $n$ twice larger and reapply all the leaf insertions that occurred so far. This way, with $n$ leaf insertion queries in total, the data structure reinitializes at most $\log n$ times, and the number of actual leaf additions over all instantiations of the data structure is bounded by $2n$.} and fix it throughout this section.
For convenience, let $V$ be the set of vertices of size $n + 1$
containing $s$ and all the vertices to be inserted to the dynamic tree.

\newcommand{\level}{\mathsf{level}}
\newcommand{\weightds}{\textcolor{blue}{TODO [use Appendix A thing]}}

On initialization, compute the constant $C \coloneqq \left\lceil 2(\gamma + \lambda) \right\rceil \geq 1$  and define the values $K \coloneqq \left\lceil C \log n \right\rceil$ and $p \coloneqq K^{-1}$; assume that $p \in (0, \frac12)$.
Let $t$ be the smallest positive integer for which $(n + 1) p^t \le 1$ (note that $t \leq \log_2(n + 1)$), and create a~finite sequence of integers $(n_0, \dots, n_t)$ by setting $n_0 = n+1$ and $n_i = \left\lceil n_0 p^i \right\rceil$.
We the define a~descending chain of hitting sets in $T$:
\[ V = L_0 \supseteq L_1 \supseteq \ldots \supseteq L_t = \{s\}, \]
by setting $L_0 = V$ and, for each $i \in \{1, \dots, t\}$, choosing $L_i$ as a~random subset of $L_{i-1}$ of size $n_i$ containing $s$.
Note that $n_t = 1$, so $L_t = \{s\}$.
For each vertex $v \in V$, define its \emph{level}, denoted $\level(v)$, as the largest integer $j \in \{0, \dots, t\}$ such that $v \in L_j$.

Observe that the interface presented in \cref{l:dist-difference-query} is deterministic: every query has exactly one correct answer (\emph{smaller}, \emph{equal} or \emph{larger}).
Therefore, at no point of time can a~user of the data structure determine any information about the internal state of the data structure, including any information about the hitting sets $L_i$ (assuming all the answers up to this point of time were correct).
Thus we can assume that the sequence of requests to the data structure (leaf insertions and queries) is predetermined, though unknown to the implementation of the data structure.
In particular, we assume that the requests eventually produce a~predetermined rooted out-tree $T^\star$.

For any $v \in V$ and $i \in \{0, \dots, t\}$, let $\alpha_v^i$, or the \emph{nearest level-$i$ ancestor} of $v$, be the nearest ancestor $w \in L_i$ of $v$ in $T^\star$.	
Note that $\alpha_v^i$ is always correctly defined as $\level(s) = t$ and $s$ is an~ancestor of every vertex of $T^\star$ (including itself).
Also note that a~data structure maintaining $T$ can maintain the values of $\alpha_v^i$ for all vertices $v$ already added to the tree.
Then we have that


\begin{lemma}
  \label{lem:all-paths-have-few-hops}
  Fix $i \in \{0, \dots, t\}$.
  Then with probability at least $1 - n^{-(k + 1)}$, for all $v \in V$, the nearest level-$i$ ancestor of $v$ is at most $\gamma \cdot K^i \log n$ hops apart from $v$ in $T^\star$.
\end{lemma}
\begin{proof}
  Observe that by the construction, $L_i$ is a~uniformly random subset of $V$ of size $n_i$ containing $s$; also note that $n_i = \left\lceil n / K^i \right\rceil$.
  Applying \cref{l:hitting_basic}, we find that with probability at least $1 - n^{-(k+1)}$, for every $v \in V$, the nearest ancestor of $v$ in $L_i$ is at most $\gamma \cdot K^i \log n$.
\end{proof}

In the following description we assume that the conclusion of \cref{lem:all-paths-have-few-hops} holds.

To continue, define the following sequences $(b_0, \dots, b_t)$, $(\ell_0, \dots, \ell_t)$ of integers:
\[
b_i = cB \cdot K^{i+1}, \qquad
\ell_i = 10 b_i \cdot K.
\]

We now sketch the implementation of the data structure.
We maintain the data structure from \cref{l:tree-query} supporting queries computing the weight of a~descending path in $T$.
Recall that the amortized update time for the data structure is $O(\polylog n)$ and the query time for a~$k$-hop path is $\Ot(k)$.
Now we aim to construct, for each $i = t, t-1, \dots, 0$, an~auxiliary data structure $\D_i$ answering queries of the form: given two vertices $u, v \in V(T)$ and a~rational number $\frac{p}{q} \in \rat[2^{b_i}]$, compare $\delta_T(s, u) - \delta_T(s, v)$ to $\frac{p}{q}$.
Note that the implementation of $\D_t$ is straightforward: since $L_t = \{s\}$, all queries reduce to the comparison of any~given rational number in $\rat[2^{b_t}]$ with $0$; this can be done easily in time $O(b_t)$.
On the opposite end of the spectrum, an~efficient implementation of $\D_0$ will directly resolve \cref{l:dist-difference-query}.

The following lemma proves that the auxiliary data structures $\D_{t-1}, \dots, \D_0$ can be constructed inductively:
\begin{lemma}\label{lem:dist-inductive-data-structures}
  Let $i \in \{0, \dots, t - 1\}$.
  Suppose we have access to the data structure $\D_{i+1}$ as an~oracle.
  Then there exists an~implementation of the data structure $\D_i$ so that, assuming that $\D_{i+1}$ answers all the queries correctly, with probability at least $1 - \frac{1}{n^{k+1}}$, $\D_i$ processes all vertex insertions and $q_i$ queries correctly:
  \begin{itemize}
    \item performing at most $\Ot(n_i)$ queries to $\D_{i+1}$,
    \item in time $\Ot(n + q_ib_i)$, not counting the time of the operations on $\D_{i+1}$.
  \end{itemize}
\end{lemma}
\begin{proof}
  Note that the answers to the queries that $\D_i$ is supposed to produce are uniquely determined.
  Consequently, as long as $\D_i$ does not err (which will be the case with high probability by construction),
  the answers of $\D_i$ will be uniquely determined.
  Hence, we can assume that, w.h.p., the sequence of queries to $\D_i$ does not depend on the answers
  produced by $\D_i$, or, in other words, is predetermined (but unknown to $\D_i$).

  The auxiliary data structure maintains:
  \begin{itemize}
    \item An~instance of the data structure for incremental sparse edge covers from \cref{l:incremental-nbd-cover}, initialized with an~edgeless graph $H$ on the vertex set $L_i$.

      From now on, we assume that the data structure maintains dynamically a~collection of sets $A_1, \dots, A_k$ satisfying, with high probability, the edge covering condition, the weak diameter condition, and the sparsity condition.
    \item For every vertex $v \in L_i$ already added to the tree:
      \begin{itemize}
        \item for $v \neq s$, $z_{v,i}$: the nearest \emph{strict} ancestor of $v$ in $L_i$; and $d_{v,i}$: the exact rational distance in $T$ from $z_{v,i}$ to $v$.
        \item $d_{v,i+1} \in \rat$: the exact rational distance in $T$ from $\alpha_{v,i+1}$ to $v$.
        \item $a_v \in \rat[2^{O(\ell_i)}]$: an~approximation of $\delta_T(s, v)$, satisfying $|a_v - \delta_T(s, v)| \le \frac{k}{2^{\ell_i + 2} \cdot n}$, where $k = |L_i \cap T[s \to v]|$ ($k \leq n$) is the number of vertices of $L_i$ on the unique path from $s$ to~$v$.
        Moreover, the denominator of $a_v$ is equal to $2^{\ell_i + 2} \cdot n$.
        
        Note that the approximation $a_v$ satisfies in particular $|a_v - \delta_T(s, v)| \leq 2^{-(\ell_i + 2)}$.
        The stronger approximation will only prove useful during the computation of the values $a_v$.
      \end{itemize}
  \end{itemize}
  
  On initialization of the data structure, we set $d_{s, i+1} = 0$ and $a_s = 0$.
  Moreover, the data structure will maintain the following invariant about $H$: if $uv \in E(H)$, then $\delta_T(s, u) \sim_{b_i}^{\ell_i} \delta_T(s, v)$.
  Since $\ell_i \gg b_i$, we actually infer from the weak diameter condition that for every pair of vertices $u, v$ in a~single cluster, the distances $\delta_T(s, u)$ and $\delta_T(s, v)$ are similar:
  \begin{claim}
    \label{cl:same-cluster-similar}
    Assume $u, v \in L_i$ belong to the same cluster $A_j$ in the sparse edge covering from \cref{l:incremental-nbd-cover}.
    Then
    \[\delta_T(s, u) \sim_{(1 + b_i) \cdot \lambda \log n}^{\ell_i - \lambda \log n} \delta_T(s, v). \]
  \end{claim}
  \begin{proof}
    By the satisfaction of the weak diameter condition in $A_j$, there exists a~path $v_0, v_1, \dots, v_z$ in $H$ of length at most $\lambda \log n$ with endpoints $u$ and $v$.
    By the invariant in $H$, we have
    \[ v_{j-1} \sim_{b_{i}}^{\ell_i} v_{j} \qquad \text{for every }j \in \{1, \dots, z\}. \]
    Applying \cref{l:frac-similarity-equivalence} multiple times, we get the conclusion of the lemma.
  \end{proof}
  For brevity, we set
  \begin{align*}
    b'_i &= (1 + b_i) \cdot \lambda \log n,\\
    \ell'_i &= \ell_i - \lambda \log n.
  \end{align*}
  Since $\ell'_i$ is still sufficiently large compared to $b'_i$, we conclude that the distances from $s$ to the vertices of a~single cluster $A_j$ can be ordered with respect to $\preceq_{b'_i}^{\ell'_i}$:
  \begin{claim}
    \label{cl:same-cluster-transitive}
    Assume $u, v, w \in L_i$ belong to the same cluster $A_j$.
    Then:
    \begin{itemize}
      \item it is always the case that $\delta_T(s, u) \preceq_{b'_i}^{\ell'_i} \delta_T(s, w)$ or $\delta_T(s, w) \preceq_{b'_i}^{\ell'_i} \delta_T(s, u)$,
      \item if $\delta_T(s, u) \preceq_{b'_i}^{\ell'_i} \delta_T(s, v)$ and $\delta_T(s, v) \preceq_{b'_i}^{\ell'_i} \delta_T(s, w)$, then $\delta_T(s, u) \preceq_{b'_i}^{\ell'_i} \delta_T(s, w)$.
    \end{itemize}
  \end{claim}
  \begin{proof}
    First, observe that:
    \begin{align*}
      \ell'_i &\geq 10b_i \cdot C\log n - \lambda \log n \geq 20b_i \cdot \lambda \log n - \lambda \log n \geq 19b_i \cdot \lambda \log n, \\
    b'_i &= (1 + b_i) \cdot \lambda \log n \leq 2 b_i \cdot \log n.
    \end{align*}
    Hence, $\ell'_i \geq 4b'_i + 5$.
    We infer from \cref{cl:same-cluster-similar} and the definition of $\preceq_{b'_i}^{\ell'_i}$ that for every pair of vertices $u, w \in A_j$, the values $\delta_T(s, u)$ and $\delta_T(s, w)$ are ordered by the relation of $(b'_i, \ell'_i)$-smallness.\footnote{Note that it can both happen that $\delta_T(s, u) \preceq_{b'_i}^{\ell'_i} \delta_T(s, w)$ and $\delta_T(s, w) \preceq_{b'_i}^{\ell'_i} \delta_T(s, u)$, in which case we say that these two values are $(b'_i, \ell'_i)$-equal, denoted $\delta_T(s, u) \approx_{b'_i}^{\ell'_i} \delta_T(s, w)$.}
    In turn, the transitivity follows immediately from \cref{cl:same-cluster-similar} and \cref{l:frac-smaller-transitivity}.
  \end{proof}
  
  Note that the relation defined by $\preceq_{b'_i}^{\ell'_i}$ on a~distances from $s$ to the vertices of a~cluster is actually a~\emph{total preorder}: it is transitive and total, but some distances may be equivalent wrt. $\preceq_{b'_i}^{\ell'_i}$.
  
  We can now exploit the transitivity property provided by \cref{cl:same-cluster-transitive} by showing that the linear order of values $\delta_T(s, v)$ for $v \in A_j$ can be maintained dynamically:
  \begin{claim}
    \label{cl:cluster-ordering-maintenance}
    There exists a~data structure maintaining the set of elements of a~single cluster $A_j$, supporting the following types of operations:
    \begin{itemize}
      \item Add (remove) an~element $v \in L_i$ to (from) $A_j$.
      It is guaranteed that after the update, the cluster satisfies the weak diameter property.
      The update is processed in time $\Ot(b_i)$, plus $O(\log n)$ accesses to the oracle $\D_{i + 1}$.
      \item Query: given two elements $u, v \in A_j$, returns whether $\delta_T(s, u) \preceq_{b'_i}^{\ell'_i} \delta_T(s, v)$.
      The query is processed in time $O(\log n)$.
    \end{itemize}
  \end{claim}
  \begin{proof}
    We maintain the elements of $A_j$ in a~balanced binary search tree (BBST; e.g., an~AVL tree) ordered by $\preceq_{b'_i}^{\ell'_i}$, with the groups of pairwise $(b'_i, \ell'_i)$-equal elements collapsed to a~single element of the BBST.
    In order to test the ordering of two vertices $u, v\in L_i$ while updating the BBST (e.g., during the insertion of $v\in L_i\setminus A_j$ to $A_j$), we implement the comparison $\delta_T(s, u) \preceq_{b'_i}^{\ell'_i} \delta_T(s, v)$ using the oracle $\D_{i+1}$.

    First, we find the unique rational number $\frac{p}{q} \in \rat[2^{b'_i}]$ satisfying $\left|\delta_T(s, u) - \delta_T(s, v) - \frac{p}{q} \right| \leq 2^{-\ell'_i}$,
    as follows.
    For any $\frac{p}{q}$ for which the bound holds, by the triangle inequality and $\ell'<\ell$, we have:
    \begin{align*}
      \left| a_u - a_v - \frac{p}{q} \right| &\leq \left|\delta_T(s, u) - \delta_T(s, v) - \frac{p}{q}\right| + \left|a_u - \delta_T(s, u)\right| + \left|a_v - \delta_T(s, v)\right|\\
      &\leq 2^{-\ell'_i} + 2 \cdot 2^{-(\ell_i + 2)}\\
      &\leq 2^{-(\ell'_i - 1)}.
    \end{align*}
    Hence, $\frac{p}{q}$ differs from the difference of the approximations $a_u - a_v$ by at most $2^{-(\ell'_i - 1)}$.
    Since $(\ell'_i - 1) \geq 2b' + 1$, it follows from \cref{l:frac-unique-similarity} that $\frac{p}{q}$ is indeed a unique fraction with this property.
    The fraction can be computed in time $\Ot(b'_i) = \Ot(b_i)$ by finding the \emph{best $b'_i$-bit rational approximation} (\cref{l:bra}), i.e., the pair $\bra(a_u - a_v, b'_i) = \left(\frac{p_1}{q_1}, \frac{p_2}{q_2}\right)$ with $\frac{p_i}{q_i} \in \rat[2^{b'_i}]$, and observing that $\frac{p}{q}$ is the fraction from $\{\frac{p_1}{q_1}, \frac{p_2}{q_2}\}$ that is closer to $a_u - a_v$.
    
    With the fraction $\frac{p}{q}$ in hand, recall that $\delta_T(s, u) \preceq_{b'_i}^{\ell'_i} \delta_T(s, v)$ if and only if $\delta_T(s, u) - \delta_T(s, v) \leq \frac{p}{q}$.
    Now notice that
    \[ \delta_T(s, u) - \delta_T(s, v) - \frac{p}{q} \ =\  \delta_T(s, \alpha_{u, i+1}) + d_{u, i+1} - \delta_T(s, \alpha_{v, i+1}) - d_{v, i+1} - \frac{p}{q}. \]
    Therefore, $\delta_T(s, u) - \delta_T(s, v) \leq \frac{p}{q}$ if and only if
    \begin{equation}
    \label{eq:distance-rewriting}
    \delta_T(s, \alpha_{u, i+1}) - \delta_T(s, \alpha_{v, i+1})\ \leq\ \frac{p}{q} + d_{v, i+1} - d_{u, i+1}.
    \end{equation}
    Recall that $\alpha_{u, i+1}, \alpha_{v, i+1} \in L_{i+1}$.
    Moreover, by \cref{lem:all-paths-have-few-hops} and the fact that the edge weights in~$T$ are in $\rat^{(c)}$, we have that $d_{u, i+1}, d_{v, i+1} \in \rat^{(c \cdot \gamma K^{i+1} \log n)}$, so $d_{v, i+1} - d_{u, i+1} \in \rat^{(2c \cdot \gamma K^{i+1} \log n)} = \rat[2^{2cB \cdot \gamma K^{i+1} \log n - 1}]$.
    Thus, the right-hand side of \cref{eq:distance-rewriting} is in $\rat[2^z]$, where
    \begin{flalign*}
    && z & = 2cB(\gamma \cdot K^{i+1} \log n) + b'_i \\
    &&   & = 2cB \cdot \gamma K^{i+1} \log n + (1 + b_i) \cdot \lambda \log n && (b'_i = (1+b_i) \cdot \lambda \log n) \\
    &&   & \leq 2cB \cdot \gamma K^{i+1} \log n + 2b_i \cdot \lambda \log n && (b_i \geq 1) \\
    &&   & \leq 2b_i \cdot \gamma \log n + 2b_i \cdot \lambda \log n && (b_i = cB \cdot K^{i+1}) \\
    &&   & = b_i \cdot 2 \log n \cdot (\gamma + \lambda) \leq b_i \cdot K = b_{i+1}.
    \end{flalign*}
    Hence, $\frac{p}{q} + d_{v, i+1} - d_{u, i+1} \in \rat[2^{b_{i+1}}]$, so $\D_{i+1}$ can be queried to compare $\delta_T(s, \alpha_{u, i+1}) - \delta_T(s, \alpha_{v, i+1})$ to $\frac{p}{q} + d_{v, i+1} - d_{u, i+1}$.
    The result of this comparison is exactly the result of the comparison between $\delta_T(s, u) - \delta_T(s, v)$ with $\frac{p}{q}$.
    All in all, this comparison can be implemented in time $\Ot(b_i)$ and issues exactly one query to $\D_{i+1}$.
    
    Summing up, an~update to $A_j$ requires adding or removing an~element in the BBST.
    This can be done in time $O(\log n)$, plus $O(\log n)$ comparisons between the elements of the BBST.
    By the discussion above, we conclude that a~single modification of the BBST can be performed in time $\Ot(b_i)$ with additional $O(\log n)$ queries to $\D_{i+1}$.
    
    With the BBST in place, the query is straightforward: given two elements $u, v \in A_j$, we have $\delta_T(s, u) \preceq_{b'_i}^{\ell'_i} \delta_T(s, v)$ if either $u$ and $v$ are in the same group of $(b'_i, \ell'_i)$-equal elements, or $u$ is earlier in the BBST ordering than $v$.
    Both conditions can be verified in time $O(\log n)$.
  \end{proof}
  
  \medskip
  We are now ready to implement the data structure.
  For each cluster $A_j$ of the sparse edge cover from \cref{l:incremental-nbd-cover}, we have an~instance of the data structure from \cref{cl:cluster-ordering-maintenance} maintaining the total preorder on $\{\dist_T(s, v)\,\colon\,v \in A_j\}$.
  We move to the implementation of the leaf insertion and the query.
  
  Suppose a~vertex $v \in L_i$ is added to $T$ as a~child of an~already existing vertex $u \in V(T)$.
  We compute the vertices $\alpha_{v, i}$ and $\alpha_{v, i+1}$, and using the data structure from \cref{l:tree-query}, we compute the exact distances $d_{v,i}$ and $d_{v,i+1}$.
  Recall from \cref{lem:all-paths-have-few-hops} that the paths from $z_{v,i}$ to $v$, and from $\alpha_{v,i+1}$ to $v$ have at most $O(K^i \log n)$ and $O(K^{i+1} \log n)$ hops, respectively; therefore, ${d_{v,i} \in \rat[2^{O(K^i \log^2 n)}]}$, $d_{v,i+1} \in \rat[2^{O(K^{i+1} \log^2 n)}]$. Both values can be computed exactly in time $\Ot(b_i)$ since $b_i = O(\log n) \cdot K^i$.

  Then, $a_v$ is computed by rounding $d_{v,i}$ down to the nearest fraction of the form $\frac{p}{2^{\ell_i+2} \cdot n}$ (so~$p$ is equal to $\left\lfloor 2^{\ell_i + 2} \cdot n \cdot d_{v,i} \right\rfloor$) and letting $a_v = a_{z_{v,i}} + \frac{p}{2^{\ell_i + 2} \cdot n}$.
  To see that this satisfies the required properties of $a_v$, let $k = |L_i \cap T[s \to v]|$. Then $|L_i \cap T[s \to z_{v,i}]| = k-1$ holds, so \linebreak
  $|a_{z_{v,i}} - \delta_T(s, z_{v,i})| \leq \frac{k-1}{2^{\ell_i + 2} \cdot n}$.
  Since $|\delta_T(z_{v, i}, v) - \frac{p}{2^{\ell_i + 2} \cdot n}| =|d_{v,i}-\frac{p}{2^{\ell_i + 2} \cdot n}|\leq \frac{1}{2^{\ell_i + 2} \cdot n}$ and \linebreak
  $\delta_T(s, v) = \delta_T(s, z_{v, i}) + \delta_T(z_{v, i}, v)$, it follows that $|a_v - \delta_T(s, v)| \leq \frac{k}{2^{\ell_i + 2} \cdot n}$.
  Observe that $a_v$ can also be computed in time $\Ot(b_i)$.
  
  \medskip
  
  Now consider handling a query.
  In the query, $\D_i$ receives two vertices $u, v \in L_i$ already added to a~tree, and a~rational number $\frac{p}{q} \in \rat[2^{b_i}]$, and $\D_i$ is to return the result of the comparison between $\delta_T(s, u) - \delta_T(s, v)$ and $\frac{p}{q}$.
  Our aim is to first try to perform the comparison using the approximations of $\delta_T(s, u)$ and $\delta_T(s, v)$, and -- if the result of the comparison is inconclusive -- use the machinery of sparse edge covers and \cref{cl:cluster-ordering-maintenance} to infer the result of the comparison, potentially asking some additional queries to $\D_{i+1}$.
  
  Compute $a \coloneqq a_u - a_v$.
  We consider three cases, depending on the value of $a$:
  \begin{itemize}
    \item Suppose that $a > \frac{p}{q} + \frac{1}{2^{\ell_i + 1}}$.
      Then from $|a_u - \delta_T(s, u)| \leq 2^{-(\ell_i + 2)}$ and $|a_v - \delta_T(s, v)| \leq 2^{-(\ell_i + 2)}$ we conclude that $\delta_T(s, u) - \delta_T(s, v) > \frac{p}{q}$, so we can return \emph{greater}.
      Note that this comparison can be performed in time $\Ot(b_i)$.
    
    \item Analogously, if $a < \frac{p}{q} - \frac{1}{2^{\ell_i + 1}}$, then we can return \emph{smaller}.
    
    \item If neither of the equalities above holds, then $|a - \frac{p}{q}| \leq 2^{-(\ell_i + 1)}$. Since $|a_u - \delta_T(s, u)| \leq 2^{-(\ell_i + 2)}$ and $|a_v - \delta_T(s, v)| \leq 2^{-(\ell_i + 2)}$, we find that $|\delta_T(s, u) - \delta_T(s, v) - \frac{p}{q}| \leq 2^{-\ell_i}$.
    As $\frac{p}{q} \in \rat[2^{b_i}]$, we infer that $\delta_T(s, u) \sim_{b_i}^{\ell_i} \delta_T(s, v)$.
    Therefore, still preserving the invariant on $H$, we can add an~edge $uv$ to $H$.
    The edge addition causes a~sequence of updates to the covering $L_i = A_1 \cup \dots \cup A_k$; each update (adding or removing a~vertex from a~set) is processed by the data structure of \cref{cl:cluster-ordering-maintenance}.
    
    Afterwards, we have $uv \in E(H)$, so by the edge covering condition of \cref{l:incremental-nbd-cover} there exists a~set $A_j$ such that $u, v \in A_j$.
    In order to locate this set, we exploit the sparsity condition: since at most $O(\log n)$ sets contain a~vertex $u$, we enumerate all the sets containing $u$ and test each of the for the containment of $v$.
    Having located the set $A_j$, we perform a~query on the corresponding instance of the data structure from \cref{cl:cluster-ordering-maintenance} and, depending on the result of the query, return \emph{smaller}, \emph{equal}, or \emph{larger}.
  \end{itemize}
  
  Observe that the sequence of queries to the sparse edge covering data structure of \cref{l:incremental-nbd-cover} is uniquely determined from the sequence of queries to $\D_i$, and the number of updates to \cref{l:incremental-nbd-cover} does not exceed the number of queries to $\D_i$.
  Since the queries to $\D_i$ are predetermined, then so are the queries to the sparse edge covering data structure.
  Hence, with probability at least $1 - \frac{1}{n^{k+1}}$, the sparse edge covering data structure processes all $q_i$ queries correctly in time $\Ot(n_i + q_i)$, and updates the covering at most $\Ot(n_i)$ times.
  Each update of the covering is in turn processed by the data structure of \cref{cl:cluster-ordering-maintenance} in time $\Ot(b_i)$, plus $O(\log n)$ queries to $\D_{i+1}$.
  Therefore, the total time required to process all the updates of the covering is $\Ot(n_ib_i) = \Ot(n)$, plus $\Ot(n_i)$ queries to $\D_{i+1}$.
  
  Finally, observe that each leaf insertion and each query is processed in time $\Ot(b_i)$, not counting the time of updates to the sparse edge covering data structure or queries to $\D_{i+1}$.
  Within this regime, all operations (insertions and queries) are processed in time $\Ot((n_i + q_i)b_i) = \Ot(n + q_ib_i)$.

  Thus, in total, any sequence of $n_i$ leaf insertions and $q_i$ queries can be processed in time $\Ot(n + q_ib_i)$, plus $\Ot(n_i)$ queries to $\D_{i+1}$.
  This concludes the proof.
\end{proof}

We finish the construction of the data structure by constructing, for each $i = t - 1, \dots, 0$, the auxiliary data structure $\D_i$ that uses $\D_{i+1}$ as an~oracle.
We claim that $\D_0$ is exactly the sought data structure for distance comparisons.

First, we estimate the success probability guarantees.
Recall that the conclusion \cref{lem:all-paths-have-few-hops} holds with probability at least $1 - \frac{1}{n^{k + 1}}$; and given that \cref{lem:all-paths-have-few-hops} holds, each auxiliary data structure $\D_0, \dots, \D_{t-1}$ processes all operations correctly also with probability at least $1 - \frac{1}{n^{k + 1}}$ each.
All the remaining elements of the data structure are deterministic.
Since $t \leq O(\log n)$, we conclude from the union bound that the probability that the entire data structure processes all operations correctly is at least $1 - \frac{O(\log n)}{n^{k + 1}} \geq 1 - \frac{1}{n^k}$, given that $n$ is large enough.

Now we show the runtime guarantees, assuming the correctness of the data structure.
The preprocessing step can be easily implemented in time $\Ot(n)$.
It remains to estimate, for each $i \in \{0, \dots, t\}$, the time required by $\D_i$ to process all queries, excluding the time of the operations by the remaining auxiliary data structures.
Let $q_0, \dots, q_t$ be the sequence of integers denoting, for each $q_i$, the number of queries to $\D_i$.
We have $q_0 = q$ and, by \cref{lem:dist-inductive-data-structures}, $q_i = \Ot(n_{i-1})$ for $i \in \{1, \dots, t\}$.
Again by \cref{lem:dist-inductive-data-structures}, $\D_i$ processes all $q_i$ queries in time $\Ot(n + q_ib_i)$ for $i \in \{0, \dots, t-1\}$.
On the other hand, the implementation of $\D_t$ is trivial and each query in worst-case time $\Ot(b_t)$; hence, $\D_t$ processes $q_t$ queries in time $\Ot(q_tb_t)$.

Therefore, the total time required in the worst case by $\D_0$ (excluding the time spent in other data structures) is $\Ot(n + q)$; and for each $i \in \{1, \dots, t\}$, the worst-case total time requirement (again excluding the time complexity of other auxiliary data structures) is bounded by
\[ \Ot(n + q_tb_t) = \Ot(n + (cB \cdot K^{i+1}) \cdot (n / K^{i-1})) = \Ot(n + n \cdot cBK^2) = \Ot(n),  \]
since $c$ is a~constant and $B, K \in O(\log n)$.
Hence, with high probability, the presented data structure processes all queries correctly in time $\Ot(n + q)$.
This finishes the proof of \cref{l:dist-difference-query} and implies \cref{t:sssp}.

%% file: scaling.tex
\section{Arbitrary precision scaling}\label{s:scaling}

In this section we show how to obtain an $\eps$-feasible price function (see Definition~\ref{d:epsfeasible}) of $G$ using a negative-weight SSSP algorithm for handling only integer-weighted graphs.
Formally, we prove:
\begin{lemma}\label{l:scaling}
    Suppose there is an algorithm running in time $T(N,M,C)=\Omega(N+M)$ that takes as input a digraph $H$ with $N$ vertices, $M$ edges, and integer edge weights of absolute value $O(C)$ and either detects a negative cycle in $H$ or outputs single-source distances in $H$ from a chosen vertex.

    Let $G=(V,E)$ be a digraph with short rational weights with absolute values of numerator and denominator bounded by $W$. Then, for any integer $k\geq 0$, one can either detect a negative cycle in $G$ or compute a $2^{-k}$-feasible
    price function $p^*$ of~$G$
    such that $p^*(v)\in \rat[O(nW\cdot 2^k)]$ in $O(T(n,m,W)+T(n,m,n)\cdot k)$ time.
\end{lemma}

In a standard way, without loss of generality assume that $G$ is augmented by adding a source~$s^*$ with $0$-weight edges to all $v\in V$. Such augmentation does not influence the shortest paths nor the existence of a negative cycle in $G$.

First of all, note that for a given short rational $q$, $|q|\leq W$, one can generate the subsequent digits of the binary expansion
\begin{equation*}
\pm b_\ell b_{\ell-1}\ldots b_0.b_{-1}b_{-2}\ldots
\end{equation*}
of $q$ (where $\ell=O(\log{W})$) in $O(1)$ time per digit using the well-known long division algorithm.
For $j\geq 0$, denote by $q^{(j)}$
the rational number \begin{equation*}
    q^{(j)}=\pm b_\ell b_{\ell-1}\ldots b_0.b_{-1}\ldots b_{-j},
\end{equation*}
that is, $q$ with the trailing bits $b_{-(j+1)},\ldots$ of the expansion truncated (or, equivalently, replaced by zeros). Observe that if $q\geq 0$ then $q-2^{-j}\leq q^{(j)}\leq q$ and otherwise
$q\leq q^{(j)}\leq q+2^{-j}$.

Denote by $\wei^{(j)}$ an \emph{integral} weight function on $E$ such that $\wei^{(j)}(e):=2^j\cdot (\wei(e))^{(j)}+1$.
Note that we have:
\begin{equation}\label{eq:weij}
    2^j\cdot \wei(e)\leq \wei^{(j)}(e)\leq 2^j\cdot \wei(e)+2,
\end{equation}
\begin{equation}\label{eq:wjub}
    \wei^{(j+1)}(e)\leq 2^{j+1}\cdot ((\wei(e))^{(j)}+2^{-(j+1)})+1=2(\wei^{(j)}(e)-1)+2=2\cdot \wei^{(j)}(e),
\end{equation}
\begin{equation}\label{eq:wjlb}
    \wei^{(j+1)}(e)\geq 2^{j+1}\cdot ((\wei(e))^{(j)}-2^{-(j+1)})+1=2(\wei^{(j)}(e)-1)=2\cdot \wei^{(j)}(e)-2.
\end{equation}
More generally, via repeated application of~\eqref{eq:wjlb}~and~\eqref{eq:wjub}, for any $0\leq j< l$ we have:
\begin{equation}\label{eq:wgen}
    2^{l-j}\cdot \wei^{(j)}(e)-2^{l-j+1}\leq \wei^{(l)}(e)\leq 2^{l-j}\cdot \wei^{(j)}(e).
\end{equation}

For $j\geq 0$, let $G^{(j)}$ be the graph $G$ with edge weights given by $\wei^{(j)}$.
More generally, for a price function $q$, denote by $G^{(j)}_q$ the graph $G^{(j)}$ with edge weights reduced by $q$. By~\eqref{eq:weij} we obtain that for any price function $q$, if $G^{(j)}_q$ has a negative cycle, then so does $G$.

Set $p_{-1}\equiv 0$. We will iteratively construct feasible price functions $p_0,\ldots,p_{k+1}$ for the graphs
\begin{equation*}
G^{(0)}_{2\cdot p_{0}^*},\ldots,G^{(k+1)}_{2\cdot p^*_{k+1}},
\end{equation*}
where $p^*_j$ (for $j\geq 1$; we set $p^*_{0}\equiv 0$) is given by
\begin{equation*}
  p^*_j(v):=\sum_{i=0}^{j-1}2^{j-i-1}\cdot p_i(v)=2p_{j-1}^*(v)+p_{j-1}(v),
\end{equation*}
or
possibly terminate with a negative cycle if we detect one in these graphs on the way. The construction will guarantee that each $p_i$ will
have \emph{integer values of absolute value $O(nW)$}.

Eventually, we will use 
\begin{equation*}
p^*:=\frac{1}{2^{k+1}}\cdot p_{k+2}^*=\frac{1}{2^{k+1}}\sum_{i=0}^{k+1}p_i\cdot 2^{(k+1)-i}=\sum_{i=0}^{k+1}\frac{p_i}{2^i}.
\end{equation*}
as our desired price function.
Observe that then, by $p_i(v)\in \mathbb{Z}$ and $|p_i(v)|=O(nW)$, we indeed have $p^*\in V\to \rat[O(nW \cdot 2^k)]$. Moreover, $p^*_{k+2}=2p^*_{k+1}+p_{k+1}$ -- being a feasible price function of $G^{(k+1)}$, satisfies for each $e\in E$:
\begin{equation*}
  2^{k+1}\cdot \wei(e)+2+p^*_{k+2}(u)-p^*_{k+2}(v)\geq w^{(k+1)}(e)+p^*_{k+2}(u)-p^*_{k+2}(v)\geq 0,
\end{equation*}
so
\begin{equation*}
  \wei(e)+\frac{p^*_{k+2}(u)}{2^{k+1}}-\frac{p^*_{k+2}(v)}{2^{k+1}}\geq \frac{-2}{2^{k+1}}=-2^{-k},
\end{equation*}
and thus $p^*=p^*_{k+2}/2^{k+1}$ is $2^{-k}$-feasible price function of $G$.

Let us now describe how the subsequent price functions $p_0,\ldots,p_{k+1}$ are constructed.
Since the graph $G^{(0)}$ has integer weights in $[-W,W+2]$, in $T(n,m,W)$ time one can obtain either a negative cycle or the price function $p_0$ equal to the single-source distances from $s^*$ in $G^{(0)}$.

For $j\geq 1$, the price function $p_j$ will be obtained by computing single-source
distances from $s^*$ in a certain subgraph \begin{equation*}
    H^{(j)}\subseteq G^{(j)}
\end{equation*}
with the price function $2\cdot p_j^*=2\cdot (2p_{j-1}^*+p_j)$ applied, such that:
\begin{enumerate}
    \item For $j\geq 1$, $H^{(j)}_{2\cdot p^*_j}$ has (integer) weights in $[-2,4n]$ (even though $G^{(j)}$
may have integer weights of magnitude $2^j$), whereas
for $j=0$, the integer edge weights lie in $[-W,W+2]$.
    \item No edge of $G^{(j)}\setminus H^{(j)}$ lies on either a negative cycle or a shortest path from $s^*$ in any of the graphs $G^{(j)},\ldots,G^{(k+1)}$.
\end{enumerate}

More specifically, let us put $H^{(0)}=G^{(0)}$. Then, $H^{(0)}$ satisfies both items above.

For $j\geq 1$, assuming the computation of $p_{j-1}$ succeeded and no negative cycle has been detected so far,  $H^{(j)}$ is obtained from $H^{(j-1)}\subseteq G^{(j-1)}$ 
as follows.
First, by~\eqref{eq:wjub}~and~\eqref{eq:wjlb} and $p_j^*=2\cdot p_{j-1}^*+p_{j-1}$, note that the edge weights of
$G^{(j)}_{2\cdot p^*_j}$ are obtained from the respective edge weights
of the non-negatively weighted $G^{(j-1)}_{p^*_j}=G^{(j-1)}_{2p^*_{j-1}+p_{j-1}}$ by multiplying by $2$ and adding an integer in range $[-2,0]$.
Moreover, by the construction of $p_{j-1}$, all distances from $s^*$ in $G_{p^*_j}^{(j-1)}$ (and thus also in $H_{p^*_j}^{(j-1)}$ are precisely $0$; recall that $G^{(j-1)}$ does not contain a negative cycle by assumption).
Therefore, unless $G^{(j)}$ contains a negative cycle, the distances from $s^*$ in $G^{(j)}_{2\cdot p^*_j}$ are within range $[-2n,0]$. 

Moreover, by the inductive assumption (item 2), $H^{(j-1)}$ contains all edges lying on negative cycles or shortest paths from $s^*$ in $G^{(j)},\ldots,G^{(k+1)}$,
and applying a price function does not change the structure of shortest paths or negative cycles,
the edge weights in the graph
\begin{equation*}
H'=\left(G^{(j)}\cap H^{(j-1)}\right)_{2\cdot p^*_j}
\end{equation*}
(containing an edge $e$ of weight $\wei^{(j)}(e)$ for all $e\in E(H^{(j-1}))$)
are integers $\geq -2$
and (unless $G^{(j)}$ contains a negative cycle) the distances from $s^*$ are within range $[-2n,0]$.
Additionally, by the inductive assumption (item 1) and the choice of $p_{j-1}$, the positive edge weights in $H'$ are of order $O(nW)$ and thus fit in $O(1)$ machine words.
Finally, we define $H^{(j)}$ to be $H'$ pruned of edges \emph{with weight higher than $4n$}.

Item 1 is satisfied for $H^{(j)}$ trivially.
In order to prove the inductive construction correct, we only need to argue that the pruned edges do not lie on either a simple negative cycle of a simple shortest path from $s^*$ in $G^{(j)},\ldots,G^{(k+1)}$.
From that it will follow that one can find a negative cycle in $G^{(j)}_{2p_j^*}$ or compute the distances from $s^*$ in that graph (that is, the integer price function $p_j$) in $O(T(n,m,n))$ time, as desired, by running the assumed algorithm on $H^{(j)}$.

By~\eqref{eq:wgen}, a (simple) shortest $s^*\to t$ path $P$ in $G^{(i)}$ ($j\leq i\leq k+1$), preserved in $H'$ by assumption on $H^{(j-1)}$, has cost at most $2^{i-(j-1)}\cdot 0=0$ wrt. price function $q:=2^{i-j}\cdot 2\cdot p^*_j$.
Similarly a negative cycle in $G^{(i)}$ (preserved in $H'$ by assumption)
has negative weight wrt. $q$.

Again by~\eqref{eq:wgen}, the reduced weight $\wei_{q}(e)$ wrt. $q$ of any edge $e\in E(H')$ in $G^{(i)}$ satisfies
\begin{equation*}
  \wei_q(e)\geq 2^{i-j}\cdot (-2)-2^{i-j+1}=2^{i-j}\cdot (-4).
\end{equation*}
On the other hand, an edge $e'\in E(H')$ with $\wei_{2\cdot p_j^*}(e')>4n$
satisfies
\begin{equation*}
  \wei_{q}(e')>2^{i-j}\cdot 4n-2^{i-j+1}=2^{i-j}\cdot (4n-2)
\end{equation*}
by~\eqref{eq:wgen}.
If $e'$ lied on either a simple negative cycle $P$ or a simple shortest path $P$ from $s^*$ in $G^{(i)}$, then we would have
\begin{equation*}
    \wei_q(P)\geq 2^{i-j}\cdot [(4n-2)+(n-1)\cdot (-4)]=2^{i-j}\cdot 2\geq 2,
\end{equation*}
which contradicts $w_q(P)\leq 0$. This concludes the proof.

Note that the graphs $H_{2\cdot p_j^*}^{(j)}$ that constitute the subsequent input
of the assumed negative-weights SSSP algorithm are obtained from the respective
predecessors via $O(m)$ arithmetic operations on integers of magnitude $O(nW)$.

Finally, computing the value $p^*(v)$ for some $v\in V$ out of $p_0(v),\ldots,p_{k+1}(v)$
naively would require $\widetilde{\Theta}(k^2)$ time, but adding up
the numbers $\frac{p_i(v)}{2^i}$ can be sped up using the identity
\begin{equation*}
    p_i^*(v)=\left(\sum_{i=0}^{\lceil k/2\rceil}\frac{p_i(v)}{2^i}\right)+\frac{1}{2^{\lceil k/2\rceil+1}}\left(\sum_{i=0}^{k-\lceil k/2\rceil}\frac{p_{i+(\lceil k/2\rceil+1)}}{2^i}\right).
\end{equation*}
Computing $p^*_i(v)$ recursively according to the above formula takes $\Ot(k)$ time.

By plugging the best-known integer negatively-weighted SSSP algorithms~\cite{BernsteinNW22, abs-2304-05279} into Lemma~\ref{l:scaling}, we
obtain:
\begin{corollary}\label{c:scaling}
    Let $G=(V,E)$ be a digraph with short rational weights. Then, for any integer $c\geq 0$, one can either detect a negative cycle in $G$ or compute a $2^{-c}$-feasible
    price function $p^*$ of~$G$
    such that $p^*(v)\in \rat[O(n\cdot 2^c)]$ in $\Ot((n+m)\cdot c)$ time.
    The algorithm is Monte Carlo randomized and the answer produced is correct w.h.p.
\end{corollary}

%% file: neg-cycle.tex
In this section we show the following theorem:

\begin{theorem}[Restatement of \cref{t:negsssp}]\label{thm:neg-sssp}
  Let $G$ be a directed graph with edge weights from~$\Qq[1]$.  Let $s \in V(G)$. There is an algorithm that either detects a negative cycle in $G$ or declares there is none and computes the single-source distances and a shortest paths tree from $s$. The algorithm runs in $\Ot(n^{\frac52})$ time. The algorithm is Monte Carlo randomized and produces correct output w.h.p.
\end{theorem}

\subsection{Cut Dijkstra}
The centrepiece of this theorem is the following algorithm, which we call the \emph{cut Dijkstra}.

\begin{lemma}\label{lem:kankan}
  Let $k$ be a positive integer, $G$ be a directed graph with edge weights given by \linebreak ${w : E(G) \to \Qq[1]}$ and no negative cycles.
	There exists a data structure that can be constructed in $\Ot(n^2k)$ time and supporting the following queries.

  Given a~vertex $s \in V(G)$, compute a function $\tilde{d} : V(G) \to \Qq[k+1] \cup \{\infty\}$ such that for all $v\in V$, $\tilde{d}(v)\geq \dist_G(s,v)$ and if $\dist_G^k(s,v)=\dist_G(s,v)$ then $\tilde{d}(v)$ is equal to $\dist_G(s, v)$.
  Apart from $\tilde{d}(v)$, for each $v$ such that $\tilde{d}(v) \neq \infty$,
  a path from $s$ to $v$ of weight $\tilde{d}(v)$ is returned. The query time is $\Ot(n^2 + n^{\frac32} k)$. The algorithm is randomized and produces a correct output w.h.p.
\end{lemma}
Note that in Lemma~\ref{lem:kankan} we do not guarantee that $\tilde{d}(v)$ is the length of a shortest path from~$s$ to $v$ consisting of at most $k$ edges. 

Before proceeding, we introduce more notation. For a vertex $v$, by $E^+(v)$ we denote the set of outgoing edges of $v$. By $N^+(v)$ we denote the sets out-neighbors of $v$. For $W \subseteq V(G)$, we denote $\bigcup_{w \in W} E^+(w)$ by $E^+(W)$.
 
A source of a digraph is any vertex with out-degree zero (a digraph could have multiple sources). A \emph{shortest paths digraph} from $s$ in $G$ (if $G$ has no negative cycles) is a digraph containing all edges $e \in E(G)$ such that $e$ lies on some shortest walk from $s$ to another vertex of $G$. If $G$ does not have zero cycles then this digraph is in fact a DAG.

	Let us now present the overview of cut Dijkstra.
  Answering a query will be done through a variant of Dijkstra's algorithm. 
	In order for Dijkstra's algorithm to work on a graph with negative edges, in the preprocessing stage we will compute an $\eps$-feasible price function given by \cref{c:scaling} for some sufficiently small $\eps$. As we will be using an $\eps$-feasible function, as opposed to a $0$-feasible function (as in Johnson's algorithm \cite{Johnson}), we will need to justify why it is sufficient for our needs. What is more, we will not be interested in shortest paths that consist of more than $k$ edges and, as a consequence, all distances that we will be compute will belong to $\Qq[k+1]$.

  The bottleneck in the naive implementation of this approach would be that we would make a priori $\Theta(n^2)$ relaxations and heap updates on elements of $\Qq[\Oh(k)]$ that could take up to $\Oh(k)$ time each, resulting in a $\Ot(n^2 k)$ time complexity.
  To avoid that, for relaxations we will rely on \cref{l:dist-difference-query} and process them in $\Ot(1)$ time. For heap updates, we will be able to perform them in a lazy manner. Using the game described in \cref{lem:game}, we will bound the total number of heap updates by~$\Oh(n^{\frac32})$, resulting in a total $\Ot(n^2 + n^{\frac32}k)$ time complexity per query.

\paragraph{Preprocessing.} Let us start with the description of the preprocessing stage. As its first step, we apply \cref{c:scaling} with $c=(2k+1)B + \ceil{\log_2 n}$ and hence obtain an $\eps$-feasible price function $p : V(G) \to \Q[\Oh(n^2 \cdot 2^{(2k+1)B})]$, where $\eps < \frac{2^{-(2k+1)B}}{n}$. This computation takes $\Ot(n^2k)$ time.

As the second preprocessing step, for each pair $u, v \in V(G)$, we compute the best rational approximation $\bra(p(u) - p(v), 2B)$ using \cref{l:bra}. Since $p(u), p(v) \in \Q[\Oh(n^2 \cdot 2^{(2k+1)B})]$, we have \linebreak
$p(u) - p(v) \in \Q[\Oh(n^4 \cdot 2^{(4k+2)B})]$. Computing such a rational approximation takes $\Ot(k)$ time. As we compute one for each pair of $u, v \in V(G)$, this takes $\Ot(n^2k)$ time as well.
	
	\newcommand{\ShouldProcess}{\mathsf{should\_process}(v)}
	
\paragraph{A generic variant of Dijkstra's algorithm.} As we will need to introduce a few changes to the classical version of Dijkstra's algorithm, let us take a higher-level look at it and show a series of simple but general properties before we go into more specific details.
  Let us consider a generalized shortest paths estimation algorithm (\cref{alg}) whose special case is the classical version of Dijkstra's algorithm.
	
  \begin{algorithm}\label{a:generic-dijkstra}
		\begin{algorithmic}
			
			\caption{Generic variant of Dijkstra's algorithm} \label{alg}
			\For {$v \in V(G)$}
			\State $\tilde{d}(v) \gets \infty$
			\EndFor
			\State $\tilde{d}(s) \gets 0$
			\State $U \gets V(G)$
			\While {$U \neq \emptyset$}
			\State choose $v \in U$ 
			\If {$\ShouldProcess$}
			\For {$u \in N^+(v) \cap U$}
			\State $\tilde{d}(u) \gets \min(\tilde{d}(u), \tilde{d}(v) + w(vu))$
			\EndFor
			\EndIf
			\State $U \gets U \setminus \{v\}$
			\EndWhile
			\Return {$\tilde{d}$}
			
		\end{algorithmic}
	\end{algorithm}
	
  In the ideal scenario, after running~\cref{alg} on $G$, $\tilde{d}(v) = \dist_G(s, v)$ should hold. The two aspects of~\cref{alg} that need to be specified is the way of choosing a next vertex $v$ from $U$ and the precise $\ShouldProcess$ condition. In a classical Dijkstra's algorithm, we never skip relaxations of edges, that is, the $\ShouldProcess$ condition always evaluates to true. In our scenario, we will skip performing relaxations for some vertices, no a more intricate condition will be needed.

When a vertex $v$ is chosen from $U$, we say that it is \emph{extracted}. When we execute the following relaxations, we say that $v$ is \emph{processed}. The key idea behind Dijkstra's algorithm is that there is a way of choosing vertices from $U$
that leads to the correct result at the end if the edges' reduced costs are non-negative.

Let us call the permutation of $V(G)$ resulting from the order that we extracted vertices in will the \emph{extraction order}.
By an \emph{enhanced order} we denote a pair $(K, L)$, where $K$ is an extraction order and $L \subseteq V(G)$ is the subset of vertices that we have decided to process.
  We say that an enhanced order $(K, L)$ is \emph{good} if using it in \cref{alg} for $G$ and $s$ produces values $\tilde{d}(v)$ such that $\tilde{d}(v) = \dist_{G \setminus E^+(V(G) \setminus L)}(s, v)$. In other words, $(K,L)$ is good if~\cref{alg} if it computes correct distances in a graph $G$ with edges coming out of vertices that we have decided to not process removed. An extraction order $K$ is \emph{good} if the enhanced order $(K, V(G))$ is good.
	
	Let us mention that assuming that an enhanced order $(K, L)$ is good, from the algorithm's run we can easily recover a tree of shortest paths for $G \setminus E^+(V(G) \setminus L)$, as follows.
	For every vertex $u$ we record the last vertex $v$ whose processing caused $\tilde{d}(u)$ to drop and denote it by $\mathrm{parent}(u) = v$.  We create a graph with edges from $\mathrm{parent}(u)$ to $u$ for each $u \neq s$, and this way get an out-branching from $s$ that constitutes a shortest paths tree from $s$ for $G \setminus E^+(V(G) \setminus L)$.
	
	In the classical version of Dijkstra's algorithm, the next vertex to extract is chosen to be the vertex $v$ with the smallest value of $\tilde{d}(v)$ out of these that were not extracted yet. 
However, if we somehow knew a good extraction order through other means, then such algorithm would work correctly even for negative weights.
	
Below we study the properties one needs to assume about the used enhanced order
to ensure~\cref{alg} computes correct exact distances for a subgraph of $G$.
	\begin{Definition}
	 	Let $H$ be a digraph, $D\subseteq H$ be a subgraph of $G$, and let $(K, L)$ be an enhanced order. We say that $(K, L)$ is \emph{compliant with $D$} if $V(D) \subseteq L$ and for every $v \in V(D)$ that is not the first vertex of $V(D)$ in $K$, there exists $u \in V(D)$ such that $uv \in E(D)$ and $u$ precedes $v$ in $K$.
	\end{Definition}
Note that for any enhanced order compliant with $D$, it has to have a unique source.
	
	\begin{claim} \label{cl:compliant}
	 	Let $H$ be a digraph, $s$ be a vertex of $H$, and $S$ be a subset of $V(H)$ containing $s$. Let $D$ be the digraph of shortest paths from $s$ in $H$ and let $(K, L)$ be an enhanced order.

    If $(K, L)$ is compliant with $D[S]$, then \cref{alg} run on $H$ from $s$ with $(K, L)$ as an enhanced order correctly computes all distances from $s$ to all vertices of $S$. That is, $\tilde{d}(v) = \dist_H(s, v)$ will hold for all $v \in S$ at the time of extracting $v$ and at the end of algorithm's execution as well.
	\end{claim}
	\begin{proof}
    We note that at any point of runtime of \cref{alg}, $\tilde{d}(v) \ge \dist_H(s, v)$ always holds for all $v\in V$. Also, the values $\tilde{d}$ never grow. Hence, if at any point we have $\tilde{d}(v) \le \dist_H(s, v)$, then actually $\tilde{d}(v) = \dist_H(s, v)$ and $\tilde{d}(v)$ will not change till the end of the execution.
		
		The claim will be proven inductively. Let $(u_1, \ldots, u_{|S|})$ be the order in which vertices of $S$ appear in $K$, that is, $K$ restricted to $S$. Note that as $(K, L)$ is compliant with $D[S]$, all vertices of $S$ are processed eventually. Let us inductively prove that \cref{alg} run on $H$ with $(K, L)$ as enhanced order correctly computes all distances from $s$ to $u_1, \ldots, u_i$ before extracting $u_i$.

  $(K, L)$ being compliant with $D$ implies that $u_1=s$, so setting $\tilde{d}(s)=0$ before the while loop makes this claim true for $i=1$. Let us assume that the induction hypothesis is true for $i-1$ and we will prove it for $i$. As $(K, L)$ is compliant with $D[S]$, there exists $j < i$ such that $D[S]$ contains an edge from $u_j$ to $u_i$. The induction hypothesis for $j$ states that when $u_j$ is extracted, we have $\tilde{d}(u_j) = \dist_H(s, u_j)$. As $u_j \in S$, $u_j$ is indeed processed. Hence, while relaxing the edge $u_ju_i$, $\tilde{d}(u_i) \gets \min(\tilde{d}(u_i), \tilde{d}(u_j) + \wei(u_ju_i))$ is executed. However, at that point we already had $\tilde{d}(u_j) = \dist_H(s, u_j)$ and as $u_ju_i \in E(D)$ we in fact have $\dist_H(s, u_i) = \dist_H(s, u_j) + w(u_ju_i)$. Hence, this ensures $\tilde{d}(u_i) = \min(\tilde{d}(u_i), \dist_H(s, u_i))\leq \dist_H(s, u_i)$. The induction hypothesis for $i$ follows.
	\end{proof}
	
	We will also use the following immediate fact:
	\begin{fact} \label{f:compliant-path}
    Let $H$ be a digraph, let $s,v\in V(H)$, and let $D\subseteq H$ be the shortest paths digraph from $s$. Suppose \cref{alg} is run on $H$ with $s$ as a source. Assume that $v$ is processed and $\tilde{d}(v) = \dist_H(s, v)$ holds at the time of processing $v$. Then, there exists a path $P$ from $s$ to $v$ in $D$ such that 
all the vertices of $P$ have been extracted and processed in the order
of their appearance on $P$.
	\end{fact}
	
\paragraph{Unoptimized Cut Dijkstra.} We are now ready to specify our algorithm.
	Let $D$ be the shortest paths digraph from $s$ in $G$.
  Let $A$ be the set of vertices $v$ of $V(G)$ such that $\dist_G(s, v) \in \Qq[k]$. Denote by $S$ be the set of vertices reachable from $s$ in $D[A]$.
  Finally, let $C$ be the set of vertices $v$ such that there exists a shortest path from $s$ to $v$ using at most $k$ edges, i.e., $\dist_G^k(s,v)=\dist_G(s,v)$. As the length of any path that has at most $k$ edges belongs to $\Qq[k]$, it is easy to see that $C \subseteq S$.
	
	Consider a specific variant of \cref{alg} that is expressed using pseudocode in \cref{alg2}.
	
	\begin{algorithm}
		\begin{algorithmic}
			
			\caption{Unoptimized cut Dijkstra} \label{alg2}
			\For {$v \in V(G)$}
			\State $\tilde{d}(v) \gets \infty$
			\EndFor
			\State $\tilde{d}(s) \gets 0$
			\State $U \gets V(G)$
			\While {$U \neq \emptyset$}
			\State choose $v \in U$ minimizing $\tilde{d}(v) - p(v)$
			\If {$\tilde{d}(v) \in \Qq[k]$}
			\For {$u \in N^+(v) \cap U$}
			\State $\tilde{d}(u) \gets \min(\tilde{d}(u), \tilde{d}(v) + w(vu))$
			\EndFor
			\EndIf
			\State $U \gets U \setminus \{v\}$
			\EndWhile
			\Return {$\tilde{d}$}
			
		\end{algorithmic}
	\end{algorithm}
  In words, \cref{alg2} is a modification of \cref{alg} such that:
  \begin{itemize}
    \item the next vertex to be extracted is the one minimizing $\tilde{d}(v) - p(v)$ (ties are broken arbitrarily),
    \item after extracting $v$ from $U$, \cref{alg2} skips doing the relaxations if $\tilde{d}(v) \not\in \Qq[k]$,
    \item when processing a vertex $v$, it iterates over the out-neighbors of $v$  that have not been extracted yet, guaranteeing that $\tilde{d}(v)$ does not change till the end of execution after $v$ gets extracted.
  \end{itemize}
	
	\begin{claim} \label{cl:processing}
		Let $v \in V(G)$ and assume that, during a run of \cref{alg2}, $v$ got processed and $\tilde{d}(v) = \dist_G(s, v)$ holds at the time of processing it. Then, $v \in S$.
	\end{claim}
	\begin{proof}
		For contradiction, assume that $v \not\in S$. By \cref{f:compliant-path}, there exists a path $u_0 u_1 \ldots u_c$ in $D$ such that $u_0 = s$, $u_c = v$, for each $i=0,\ldots,c-1$, $u_i$ was extracted before $u_{i+1}$ and processed. This implies that $\tilde{d}(u_i) = \dist_G(s, u_i)$ at the time of processing $u_i$ for each $i=0, \ldots, c$. As $v \not\in S$, there exists the smallest $i$ such that $u_i \not\in S$. We know that $i \ge 1$ as $s \in S$. However, as $\tilde{d}(u_i) = \dist_G(s, u_i)$ at the time of extracting $u_i$ and $u_i$ was processed, we have $\dist_G(s, u_i) \in \Qq[k]$. But $u_{i-1} \in S$ and $\dist_G(s, u_i) \in \Qq[k]$ would imply that $u_i \in S$ based on the definition of $S$, which is a contradiction.
	\end{proof}
	
	\begin{lemma} \label{lem:kankan-specification}
		\cref{alg2} produces an enhanced order that is compliant with $D[S]$.
	\end{lemma}
	\begin{proof}
		First, let us note that skipping relaxations if $\tilde{d}(v) \not\in \Qq[k]$ guarantees that whenever we execute $\tilde{d}(u) \gets \min(\tilde{d}(u), \tilde{d}(v) + w(vu))$, we have $\tilde{d}(v) + w(vu) \in \Qq[k+1]$. Hence, at any point of runtime, for all $u \in V(G)$ we have $\tilde{d}(u) \in \Qq[k+1] \cup \{\infty\}$.
		
		Observe that $s$ is the first extracted vertex, as it is the only vertex with finite $\tilde{d}$ value at the beginning. Moreover, $s$ is clearly processed.
		
		Let us call a vertex $v \in S$ a \emph{violator} if $v \neq s$ and $v$ is either not processed at all or at the moment of extracting it, there does not exist any vertex $u\in S$ such that $u$ has already been processed and $uv \in E(D[S])$.
    The existence of any violator is equivalent to the fact that the enhanced order produced by this algorithm is not compliant with $D[S]$.
		Hence, let us proceed by contradiction and assume that the set of violators is non-empty. Let $v$ be the violator that is extracted first.
		
		Let us denote by $R$ the set of vertices from $S$ that had been extracted before $v$. Let us apply \cref{cl:compliant} with the set $R$ as $S$ from its statement. One can note that as $v$ was the first violator, $(K, L)$ has to be compliant with $D[R]$, so in particular for any $u \in R$ we have that $\tilde{d}(u) = \dist_G(s, u)$ at the time of extracting $u$, so since $u \in S \Rightarrow \dist_G(s, u) \in \Qq[k] \Rightarrow \tilde{d}(u) \in \Qq[k]$, we get that $u$ was processed as a consequence.
		
		\medskip
			\textbf{Case 1:} \emph{There exists $u \in S$ such that $u$ had  already been processed before $v$ was extracted and $uv \in E(D[S])$.}
			
			\smallskip
		
		We have that $u \in R$. As $\tilde{d}(u) = \dist_G(s, u)$ at the time of processing $u$ and $uv \in E(D)$, it means that $\tilde{d}(v) = \dist_G(s, v)$ at the time of extracting $v$. As $v \in S$, we have that $\dist_G(s, v) \in \Qq[k]$, so $\tilde{d}(v) \in \Qq[k]$, but that in turn implies that $v$ was processed. That connected with the fact of $u$'s existence, we conclude that $v$ was not a violator, so we get a contradiction in this case. 
		
		\medskip		
		
		\textbf{Case 2:} \emph{There does not exist any $u \in S$ such that $u$ had already been processed before~$v$ was extracted and $uv \in E(D[S])$.}
		
		\smallskip
		
    Let us consider any path in $D[S]$ from $s$ to $v$ (it has to exist based on the definition of $S$) and let $u$ be the last vertex on that path such that $u \in R$ ($s \in R$, so $u$ is well defined). Let $u_0 u_1 \ldots u_{c-1} u_c$ be the suffix of that path such that $u_0=u$ and $u_c = v$. As $u$ was processed, we have that $c \ge 2$ based on the assumption of this case.
		
		Note that as $u \in R$, when processing $u$ it already held that $\tilde{d}(u) = \dist_G(s, u)$, so when relaxing through an edge $u_0u_1$, the value $\tilde{d}(u_1)$ was set to $\dist_G(s, u_1)$, hence at the time of processing $v$ it already holds that $\tilde{d}(u_1) = \dist_G(s, u_1)$.

		As $v$ was extracted before $u_1$, we have that at the time of $v$'s extraction we had \linebreak $\tilde{d}(v) - p(v) \le \tilde{d}(u_1) - p(u_1)$. However, as $u_1 u_2 \ldots u_{c-1} u_c$ is a path in the shortest paths digraph, we have that $\dist_G(s, v) = \dist_G(s, u_1) + w(u_1u_2) + \ldots + w(u_{c-1}u_c)$.

		Because of these, we can get the following chain of inequalities holding at the time of processing~$v$:
		
		\begin{align*}
		\dist_G(s, u_1) - p(u_1) &= \tilde{d}(u_1) - p(u_1) \\
		  &\ge \tilde{d}(u_c) - p(u_c) \\
		  &\ge \dist_G(s, u_c) - p(u_c) \\ 
		  &= \dist_G(s, u_1) - p(u_c) + \sum_{i = 1}^{c-1} w(u_i u_{i+1}) \\
		  &= \dist_G(s, u_1) - p(u_1) + \sum_{i=1}^{c-1} (w(u_i u_{i+1} + p(u_i) - p(u_{i+1})) \\
		  &\ge \dist_G(s, u_1) - p(u_1) - (c-1) \eps.
		\end{align*}
		
		
		As the left hand side and the right hand side differ by $(c-1) \eps$, we conclude that in the inequality $\tilde{d}(u_c) - p(u_c) \ge \dist_G(s, u_c) - p(u_c)$, the left hand side and right hand side cannot differ by more than $(c-1) \eps$ either. Consequently, we get that $0 \le \tilde{d}(u_c) - \dist_G(s, u_c) \le (c-1) \eps < n \cdot \eps < 2^{-((2k+1)B-1)}$. However, we have $\tilde{d}(u_c) \in \Qq[k+1]$ due to the introduced condition skipping the processing of some vertices and $\dist_G(s, u_c) \in \Qq[k]$ since $u_c \in S$. Therefore, $\tilde{d}(u_c) - \dist_G(s, u_c) \in \Qq[2k+1] = \Q[2^{(2k+1)B-1}]$. But $0$ is the only number in the interval $[0, 2^{-((2k+1)B-1)})$ belonging to $\Q[2^{(2k+1)B-1}]$, so in fact $\tilde{d}(u_c) = \dist_G(s, u_c)$ at the time of processing $u_c$. That in turn, along with \cref{f:compliant-path}, shows that there is a $t \in V(G)$ such that $t$ was processed before $v$, $\tilde{d}(t) = \dist_G(s, t)$ at the time of processing $t$ and $tv \in E(D)$. However, based on \cref{cl:processing}, this implies that $t \in S$. That in turn is a contradiction with our assumption for the current case.
	\end{proof}


	By combining \cref{lem:kankan-specification} with \cref{cl:compliant} and the fact that $C \subseteq S$, we get the following corollary.
	\begin{corollary}
    When run on $G$ with source $s$, \cref{alg2} produces values $\tilde{d}(\cdot)$  satisfying the requirements of \cref{lem:kankan}. That is, for each $v \in C$ we have $\tilde{d}(v) = \dist_G(s, v)$ at the end of its execution.
	\end{corollary}

  When implemented in the most direct way, \cref{alg2} takes $\Ot(n^2k)$ time for two different reasons. First, the line $\tilde{d}(u) \gets \min(\tilde{d}(u), \tilde{d}(v) + w(vu))$ may be executed $\Omega(n^2)$ times and since the numbers involved belong to $\Qq[k+1]$, each execution takes $\Ot(k)$ time. Second, the line ``choose $v \in U$ minimizing $\tilde{d}(v) - p(v)$'' will be executed $n$ times. If we iterate naively through the $\Oh(n)$ candidates from $U$, then the fact that each candidate is processed in $\Ot(k)$ time results in $\Ot(n^2k)$ time bound again. There is an alternative way of implementing this step used by classical version of Dijkstra on sparse graphs, i.e., maintaining the vertices to be extracted in a priority queue, e.g., a binary heap. Then, one could find the vertex minimizing $\tilde{d}(v) - p(v)$ in $\Ot(1)$ time. However, the heap would need to be updated after every relaxation. Since there is up to $\Oh(n^2)$ edge relaxations and each heap update (implemented naively) would require $\Ot(k)$ time, this would lead to an $\Oh(n^2 k)$ time bound again.
	
  \paragraph{Implementing relaxations efficiently.}
  In order to optimize relaxations time, we refer to the data structure of \cref{l:dist-difference-query}. For each extracted (but not necessarily processed) vertex $v$, we store $\tilde{d}(v)$ explicitly (and this value will not change, as explained earlier). For each $v$ yet to be extracted, we maintain $\tilde{d}(v)$ implicitly as $\tilde{d}(u) + w(uv)$ for some processed $u$. To this end, it is enough to store the corresponding vertex~$u$. Let us denote by $\mathrm{par}(v) \in V(G) \cup \{\perp\}$ a vertex $u$ that caused the last drop of $\tilde{d}(v)$ (or $\perp$ if $\tilde{d}(v) = \infty$). In other words, $\mathrm{par}(v)$ is the parent of $v$ in the shortest paths tree of the part of the graph that we have seen so far. Equipped with this, when relaxing an edge $vu$, we will compare $\tilde{d}(v) + w(vu)$ with $\tilde{d}(\mathrm{par}(u)) + w(\mathrm{par}(u)u)$ (or $\infty$ if $\mathrm{par}(u)=\perp$). Recall that $\mathrm{par}(u)$ has already been processed and $v$ is getting processed right now, which means that their $\tilde{d}(\cdot)$ values will never change. We also have $w(vu), w(\mathrm{par}(u)u) \in \Qq[1]$.
	
	Let us initialize a data structure $\D$ from \cref{l:dist-difference-query} for $c=2$. Upon extraction of a vertex $v$, it is added to $\D$ with parent $\mathrm{par}(v)$ and the parent edge weight $w(\mathrm{par}(v)v)$. As such, $\D$ will represent the tree of found shortest paths for vertices extracted so far (note that it might be different than the shortest paths tree on the graph induced on the extracted vertices, as we are skipping some relaxations). This way, for any processed vertex $v$, $\dist_T(s, v) = \tilde{d}(v)$ holds. Now, whenever we need to perform a comparison of $\tilde{d}(v) + w(vu)$ with $\tilde{d}(\mathrm{par}(u)) + w(\mathrm{par}(u)u)$, the result of this comparison is the same as the result of comparing $\tilde{d}(v) - \tilde{d}(\mathrm{par}(u))$ with $w(\mathrm{par}(u)u) - w(vu)$. Note that fhe former umber equals $\dist_T(s, v) - \dist_T(s, \mathrm{par}(u))$, while the latter belongs to $\Qq[2]$. Thus, $\D$ can perform such a comparison correctly with high probability and in $\Ot(1)$ amortized time. We conclude that the time needed for performing relaxations gets optimized to $\Ot(n^2)$. 
	
\paragraph{Implementing the priority queue.}
  We now turn to optimizing the time needed for choosing the vertex minimizing $\tilde{d}(v) - p(v)$ out of the unprocessed. 
To this end, we will show how to reduce the number of required
priority queue's (also called the heap) key updates resulting from relaxations from $\Oh(n^2)$ to $\Oh(n^{\frac32})$. 
	
  We additionally maintain auxiliary values $t(v)\in \Z_{\ge 0} \cup \{\infty\}$ satisfying the following. If $v$ has already been extracted or is on the heap, then $t(v) = \infty$. On the other hand, $t(v)\neq\infty$ means that~$v$ has not been extracted yet, it is currently not on the heap, and we can guarantee that it will not be among the next $t(v)-1$ extracted vertices (according to the chosen order) unless the value $\tilde{d}(v)$ further decreases before the next $t(v)-1$ vertices are extracted.
In particular, the above implies that at any point of time possibly only a subset
of unprocessed vertices will be present on the heap.

The values $t(\cdot)$ maintained and exploited as follows.
Initially we put $t(v)=\infty$ for all $v\in V$. Moreover, we put all the vertices $v\in V$ with the respective key $0$ (for $s$) or $\infty$ on the heap.

Before extracting a vertex, we decrease all elements of $t$ by one. For all vertices $u$ such that $t(u)$ becomes zero, we directly calculate the current value $\tilde{d}(u) - p(u)$ and put $u$ on the heap with that value as the key. Afterwards, we set $t(u)$ to $\infty$.

Then, we extract a vertex with the smallest key --- let it be~$v$. If we decide to process~$v$, we should perform all the resulting edge relaxations.

Let $R$ be the set of vertices $u$ such that $\tilde{d}(u)$ has dropped when processing $v$. We first sort all $u\in R$ wrt. $w(vu) - p(u)$ (breaking ties arbitrarily), thus getting an order $(u_1, \ldots, u_r)$, where $r = |R|$. Let us observe that\begin{equation*}
  w(vu') - p(u') < w(vu'') - p(u'') \Leftrightarrow w(vu') - w(vu'') < p(u') - p(u'')
\end{equation*}
and $w(vu') - w(vu'') \in \Qq[2] \subseteq \Q[2^{2B}]$. Recall that during the preprocessing phase we have already computed approximations $\bra(p(u') - p(u''), 2B)$. As a result, any comparison of the form \linebreak $w(vu') - p(u') < w(vu'') - p(u'')$, where $u',u''\in R$, can be performed in $\Ot(1)$ time. Since sorting a sequence of length $r$ requires $\Ot(r)$ comparisons, we can compute the order $(u_1, \ldots, u_r)$ in $\Ot(r) = \Ot(n)$ time. Observe that for each $u \in R$ we have $\tilde{d}(u) = \tilde{d}(v) + w(vu)$, so $\tilde{d}(u) - p(u) = \tilde{d}(v) + w(vu) - p(u)$. Therefore, $w(vu') - p(u') < w(vu'') - p(u'')$ implies $\tilde{d}(u') - p(u') < \tilde{d}(u'') - p(u'')$. We conclude that the obtained  order of $R$ is (at that point of time) consistent with our target extraction order given by the values $\tilde{d}(v) - p(v)$.\footnote{For safety reasons, we may prefer to break the ties in a consistent way across the whole algorithm, for example in case of equal values, preferring the one referring to the vertex with the smaller index.} 

Next, define values $h(v) \in \Z_{\ge 1} \cup \{\infty\}$ for all $v\in V$ by putting $h(u_i) = i$ for $i=1,\ldots,r$, and $h(u) = \infty$ for $u \not \in R$. We then remove from the heap all the vertices $x\in R$. Then, we set $t(u) \coloneqq \min(t(u), h(u))$ for all $u \in V(G)$. Observe that at this point, the following guarantee holds for any $i=1,\ldots,r$: unless $\tilde{d}(u_i)$ gets decreased further, the vertices $u_1, u_2, \ldots, u_{i-1}$ will be extracted earlier than $u_i$. Hence, we conclude that the invariant posed on $t$ is satisfied after the update of the values $t(\cdot)$ and the heap. 

Manipulating the values $t(\cdot)$ and $h(\cdot)$ clearly costs $O(n)$ time per extracted vertex. Hence, to bound the running time of the algorithm, it remains
to bound the total cost of the heap operations: min-extractions, insertions, and deletions of elements. If we use a standard binary heap, then each operation
costs $O(\log{n})$ times the comparison cost, which in our case is $\Ot(k)$,
as the elements' keys
are of the form $\tilde{d}(u) - p(u)\in \Q[\Oh(n^2 \cdot 2^{(3k+2)B})]$.
Since there are $n$ min-extractions in total, their cost can be bounded by $\Ot(nk)$. Since each deletion of a heap element can be charged to the corresponding
insertion of that element, it is enough bound the total number of heap insertions in the process.
To this end, note that each \emph{non-initial} insertion of a vertex $v$ on the heap can be charged to a single occurrence of the event ``the value $t(v)$ is decreased to $0$''.
We analyze the number of such events using the following abstract game.

\begin{lemma}\label{lem:game}
	Alice and Bob are playing the following game. Let $n$ be a positive integer and let $a_1, \ldots, a_n$ be a changing sequence of numbers from $\{0, 1, \ldots, n, \infty\}$. Initially, $a_i = \infty$ for all $1 \le i \le n$.  Alice and Bob perform their actions alternately, in $n$ turns as follows. First, Alice decreases each $a_i$ by one; afterwards, for each $i$ such that $a_i = 0$ she pays a dollar and sets $a_i\coloneqq \infty$. Then, Bob chooses a sequence $b_1, b_2, \ldots, b_n$ that equals $(1, 2, \ldots, c, \infty, \ldots, \infty)$ for some $1 \le c \le n$ if sorted and for each $i$, he sets $a_i \coloneqq \min(a_i, b_i)$.

  Regardless of Bob's choices, Alice will pay at most $2n \sqrt{n}$ dollars throughout the game.  
\end{lemma}
\begin{proof}
	Note that due  to how the moves are defined, after each Alice's action, all numbers will be contained in $\{1, \ldots, n, \infty\}$ and the same stays true after Bob's action. Hence, all values of $a_i$ are always contained in $\{0, 1, \ldots, n, \infty\}$ as required by the definition.
	
	For each dollar that Alice paid, let us associate it with a pair $(i, t_a)$ denoting that she paid for $a_i$ in the turn $t_a$. For each such pair, let $t_b$ be the last turn before $t_a$ when Bob changed the value of $a_i$ (as $a_i = \infty$ initially, $t_b$ is well defined). If $t_a - t_b \le \sqrt{n}$, then we call this dollar \emph{short} (that is, the time between the moment when Bob updated $a_i$ for the last time and the moment when Alice paid for it was short). Otherwise, we will call the dollar paid \emph{long}.
	
	We will prove that Alice paid at most $n \sqrt{n}$ short dollars and that she paid at most $n \sqrt{n}$ long dollars, which will clearly imply the thesis.
	
	Let us group short dollars by their value of $t_b$. Observe that for each $t_b$ there could be at most~$\sqrt{n}$ short dollars associated with it, as the value of $b_i$ used by Bob in turn $t_b$ had to be at most~$\sqrt{n}$. This implies that there is indeed at most $n \sqrt{n}$ short dollars.
	
	For long dollars, we group them by $i$. Note that for a fixed $i$, all intervals $[t_b, t_a)$ are disjoint. However, all of them are of length at least $\sqrt{n}$, whereas their sum is contained within $[0, n)$. Hence, there could be at most $\sqrt{n}$ long dollars per each $i$, implying that there are at most $n \sqrt{n}$ long dollars. This concludes the proof.
\end{proof}
If we apply~\cref{lem:game} with $a:=t$ and $b:=h$, we obtain that
the total number of heap insertions is at most $n+2n\sqrt{n}=O(n^{3/2})$.
Hence we conclude that the total cost of the heap operations is $\Ot(n^{3/2}k)$.

Summing up, we obtain that the query time of the data structure is $\Ot(n^2+n^{3/2}k)$, as desired. As we have already mentioned, maintaining the tree of paths witnessing the computed values~$\tilde{d}(v)$ can be easily done in $\Ot(n^2)$ total additional time. This finishes the proof of \cref{lem:kankan}.

\subsection{Application of cut Dijkstra}

Let us fix $G$ and $s$ from the statement of \cref{thm:neg-sssp} from that point on. Let $k \in \Z_{\ge 1}$ be a parameter. Let us assume for now that there is no negative cycle in $G$. We will get rid of this assumption at the end. 

  First, we set up the data structure given by \cref{lem:kankan} (with the chosen parameter $k$). We draw a random subset $M$ of $V(G) \setminus \{s\}$ of size $\min\left(n,\ceil{\frac{\gamma n \ln n}{k}}\right)$ for some constant $\gamma > 2$ and add the vertex $s$ into it. We will call the set $M$ as \emph{hitting set}. For each $v \in M$ we make a single query to the structure of \cref{lem:kankan} that we have just set up, thus getting a family of functions $\tilde{d}_v$, where $v \in M$. The total time of preprocessing and answering queries is $\Ot(n^2 k + \frac{n}{k} \cdot (n^2 + n^{\frac32}k)) = \Ot(n^2k + n^{\frac52} + \frac{n^3}{k})$.

  Let us now set up an auxiliary directed graph $H$ such that $V(H) = M$ and $E(G)$ forms a full bidirectional clique, where the weight of the edge $uv$ equals $\tilde{d}_u(v)$. Next, we compute single source shortest paths on $H$ from $s$ using the standard Bellman-Ford algorithm. As the edge weights of $H$ belong to $\Qq[k+1] \cup \{\infty\}$ and $H$ has size $\Ot(\frac{n}{k})$, all the intermediate values in Bellman-Ford belong to $\Qq[\Ot(n)] \cup \{\infty\}$; hence the unit cost of arithmetic operations therein does not exceed $\Ot(n)$. Applying Bellman-Ford algorithm takes $\Oh(|V(H)|^3)$ arithmetic operations. Consequently, this computation takes $\Ot((\frac{n}{k})^3 \cdot n) = \Ot(\frac{n^4}{k^3})$ time. Observe that since $\tilde{d}_u(v) \ge \dist_G(u, v)$ for all $u,v\in M$, we have $\dist_H(u, v) \ge \dist_G(u, v)$. 
  As we will show,  with high probability, this inequality is in fact an equality.

As the next step, for each vertex $v \in V(G)$ we compute the minimum value of the expression $\dist_H(s, t) + \tilde{d}_t(v)$ and the value of $t$ that minimizes it, where $t\in M$. Let us denote that minimum value as $\hat{d}(v)$. There are $n$ choices for $v$, $\Ot(\frac{n}{k})$ choices for $t$ and the arithmetic operations take $\Ot(n)$ time, hence doing so will take $\Ot(\frac{n^3}{k})$ time.

The following lemma motivates the process so far:
\begin{lemma} \label{lem:compr-paths}
With high probability,
for each $v \in V(G)$ it holds that $\hat{d}(v) = \dist_G(s, v)$
\end{lemma}
\begin{proof}
Firstly, let us note that $\dist_H(s, t) + \tilde{d}_t(v) \ge \dist_G(s, t) + \dist_G(t, v) \ge \dist_G(s, v)$, hence $\hat{d}(v) \ge \dist_G(s, v)$.	
	
For each $v \in V(G)$, fix $P_v$ to be a shortest path in~$G$ from $s$ to $v$.
  In a standard way (see Fact~\ref{l:hitting_basic}~\cite{UY91}), one can prove that with high probability (controlled by the constant $\gamma$) each $k$-edge subpath
  of any $P_v$ contains a vertex from $M$.

Fix some $v\in V$ and let $P=P_v= v_0 v_1 \ldots v_p$, where $s=v_0$ and $v=v_p$.
  Let $I = \{i_0, \ldots, i_l\}$ be the positions of the vertices from $M\cup\{v\}$
  on the path $P$, that is $V(P)\cap (M\cup\{v\})=l+1$ and $v_{i_j}\in M\cup\{v\}$. Moreover, we can assume $i_0 < i_1 < \ldots < i_l = p$, $i_0=0$ by $s\in M$ and $v_p=v$. As we have argued, for each $0 \le j < l$ we have $i_{j+1}-i_j \le k$. Note that as $P$ was the shortest path, then each of its subpaths is a shortest path between its ends as well, hence the subpath $v_{i_j} v_{i_j + 1} \ldots v_{i_{j+1}}$ is a shortest path from $v_{i_j}$ to $v_{i_{j+1}}$ whose hop-length is at most $k$. As a result, $\tilde{d}_{v_{i_j}}(v_{i_{j+1}})$ is equal to $\dist_G(v_{i_j}, v_{i_{j+1}})$ by \cref{lem:kankan}. Hence, for each $j < l-1$ the edge from $v_{i_j}$ to $v_{i_{j+1}}$ in $H$ will have its cost set to $\dist_G(v_{i_j}, v_{i_{j+1}})$. Because of this, the distance from $s$ to $v_{i_{l-1}}$ in $H$ will certainly not be bigger than $\dist_G(s, v_1) + \dist_G(v_1, v_2) + \ldots + \dist_G(v_{i_{l-2}}, v_{i_{l-1}}) = \dist_G(s, v_{i_{l-1}})$. However, as each value of $\tilde{d}_{u}(w)$ is either equal to $\infty$ or to a value of some path in $G$ from $u$ to $w$, then any path of finite cost in $H$ from $s$ to $v_{i_{l-1}}$ can be associated to some path from $s$ to $v_{i_{l-1}}$ in $G$, so its cost cannot be smaller than $\dist_G(s, v_{i_{l-1}})$. Therefore, $\dist_H(s, v_{i_{l-1}}) = \dist_G(s, v_{i_{l-1}})$. Also, since the shortest path from $v_{i_{l-1}}$ to $v$ has at most $k$ edges, then $\tilde{d}_{v_{i_{l-1}}}(v) = \dist_G(v_{i_{l-1}}, v)$. Observe that for $t=v_{i_{l-1}}$ we have that $\dist_H(s, t) + \tilde{d}_t(v) = \dist_G(s, t) + \dist_G(t, v) = \dist_G(s, v)$, hence $\hat{d}(v) \le \dist_G(s, v)$. However, we have already shown that $\hat{d}(v) \ge \dist_G(s, v)$, which implies $\hat{d}(v) = \dist_G(s, v)$.
\end{proof}

From that point on, we assume that all events that happen with high probability actually occur and thus the values $\hat{d}(v)$ that we have computed are true distances $\dist_G(s, v)$.

We now show how to find a shortest path $P_v$ from $s$ to $v$ for all $v \in V(G)$. Let us recall that the query procedure from \cref{lem:kankan} provides us with paths witnessing the values of $\tilde{d}(v)$ it returns. Let us fix $v \in V(G)$ and let $t$ be the vertex minimizing $\dist_H(s, t) + \tilde{d}_t(v)$. The classical Bellman-Ford implementation can easily return all shortest paths from the source in $\Oh(|V(H)|^3)$ time, so let $h_0 h_1 \ldots h_p$, where $h_0 = s$ and $h_p = t$ be any such returned shortest path from $s$ to $t$ in $H$. Then,
\begin{equation*}
  \dist_G(s, v) = \dist_H(s, t) + \tilde{d}_t(v) = \tilde{d}_{h_0}(h_1) + \tilde{d}_{h_1}(h_2) + \ldots + \tilde{d}_{h_{p-1}}(h_p) + \tilde{d}_t(v).
\end{equation*}
For each of these summands, we can obtain a path of the corresponding length. The concatenation of these paths will be a shortest walk from $s$ to $v$. In case it is not a simple path, i.e., it has vertex repetitions, the weight of the subpath between any two repetitions has to be zero and can be removed. After all such removals we get a simple path; let us call it $P_v$. That computation costs $\Oh(\frac{n^2}{k})$ time per vertex $v$, so computing $P_v$ for all $v \in V(G)$ takes $\Oh(\frac{n^3}{k})$ time.

To convert the collection of shortest $s\to v$ paths $P_v$ into a shortest paths tree, it is enough to find any out-branching from $s$ in the graph $D=\left(V(G),\bigcup_{v \in V(G)} P_v\right)$ (we treat paths as sets of edges here).
Indeed, each edge in this graph lies on some shortest path from $s$, and
any $s\to v$ path consisting of such edges is a shortest path in $G$.
Moreover, every $v$ reachable from $s$ in $G$ is reachable from
$s$ in $D$ as well by $P_v\subseteq D$. We conclude that a shortest paths tree
from $s$ in $G$ can be constructed in $O(n^2)$ additional time.





Gathering all the pieces, the total time complexity of this algorithm is $\Ot(n^2k + n^{\frac52} + \frac{n^3}{k} + \frac{n^4}{k^3})$. By setting $k = \ceil{\sqrt{n}}$,  we get the desired bound $\Ot(n^{\frac52})$.

\paragraph{Detecting a negative cycle.} As the final step, we explain how to get rid of the assumption that $G$ has no negative cycle. Instead of analyzing what could happen inside our algorithm if $G$ contained a negative cycle and trying to detect that at each stage, let us treat the problem in a black box way. If we consider any algorithm that works only under the assumption that its input satisfies some constraint and we run the algorithm on an input that does not satisfy the constraint anyway, then the algorithm could either crash, run for too long, or terminate and return some output (that is likely of no use). Any runtime errors can be caught externally. As we have a specific bound on the time complexity, we could also easily detect the case when it runs longer than it should. Hence, the only behavior of the algorithm run on a faulty input that is not directly distinguishable from its behavior on a valid input is when it terminates correctly and returns an erroneous output. Therefore, it suffices to find a way of validating whether the output returned by the algorithm is correct. We achieve that via the following lemma:

\begin{lemma}
	Let $T$ be an out-branching from $s$ and $d : V(G) \to \Qq[n]$ be some function.
  With high probability, one can whether $T$ is a shortest paths tree from $s$ and whether $d$ represents exact distances from $s$ in $G$ in $\Ot(n^2)$ time.
\end{lemma}
\begin{proof}
	First, we check if $d(s)=0$. Then, for each $uv \in E(T)$ we check whether $d(u) + w(uv) = d(v)$. If any of these equalities does not hold, then we return false. Checking that takes $\Ot(n^2)$ time as $T$ has $n-1$ edges and checking each equality takes $\Ot(n)$ time.
	
  If the initial check has been passed successfully, we construct the data structure from \cref{l:dist-difference-query} for $c=1$ and the out-branching $T$ as input. At this point we know that for each $v \in V(G)$, $d(v) = \dist_T(s, v)$. For such a $d$ to be the correct distance function from $s$ in $G$, it is enough that for each $uv \in E(G)$, the inequality $d(u) + w(uv) \ge d(v)$ holds. Recall that all such checks can be processed by the data structure of~\cref{l:dist-difference-query} in $\Ot(n^2)$ total time.
	
	If all the performed checks have been passed, it means that $T$ is a correct shortest paths tree and that $d$ is the correct distance function from $s$. 
  The total verification time is $\Ot(n^2)$ and the returned answer is correct with high probability. 
\end{proof}

Adding such an external check allows transforming an algorithm working only on digraphs with no negative cycles into an algorithm that works on all directed graphs and detects the existence of a negative cycle. Specifically, the algorithm declares the existence of a negative cycle if the external check fails.


%% file: appendix.tex
\section{Reporting path weights in an incremental tree}\label{s:report-distance}
For each vertex $v\in V(T)$ of the tree, let $d_v$ be the depth of $v$ in the tree.
The values $d_v$ and parent pointers can be maintained easily in $O(1)$ time per leaf
insertion.

Given a query about the weight of a descending $k$-hop path from $u$ to $v$
in $T$, we first enumerate the individual edge weights on the path $T[u\to v]$
by following $k$ parent pointers starting at $v$.

Finally, observe that a sum of $k$ short rational numbers $a_1,\ldots,a_k$ can be evaluated
in $\Ot(k)$ time, as follows.
Assume wlog. that $k$ is a power of two.
Build a full binary tree with leaves corresponding to $a_1,\ldots,a_k$.
By proceeding bottom-up, for each internal node $y$ in the tree, compute a partial sum
of all the leaves in its subtree of $y$ by adding the partial sums
stored in the children of $y$.

At the end of the process, the desired sum is stored in the root.
At the $j$-th level of the tree (counting from the bottom), the cost of computing each
of the $O(k/2^i)$ sums can be seen to be $\Ot(2^i)$.
As a result, the total cost per level is $\Ot(k)$.
Since there are $O(\log{k})$ levels, the total summation time is $\Ot(k)$.

\section{Proof of Lemma~\ref{l:bra}}\label{s:bra}
In this section, we refer to the exposition of best rational approximations
in~\cite{bra,lovasz-alt}.
Every $k$-bit rational number $\alpha$ can be expressed uniquely as a finite \emph{continued fraction}
\begin{equation*}
  \alpha=a_0+\frac{1}{a_1+\frac{1}{\ddots+\frac{\vdots}{a_{N-1}+\frac{1}{a_N}}}},
\end{equation*}
denoted also $[a_0; a_1; \ldots; a_N]$, where $a_1,\ldots,a_N$ are positive integers
and $a_0$ is an arbitrary integer. 
In fact, the numbers $a_0,\ldots,a_N$ are the successive quotients obtained
when applying the Euclidean algorithm to the numerator and denominator of $\alpha$.
Furthermore, since there exists a near-linear time implementation
of Euclid's algorithm~\cite{schonhage1971schnelle} (the naive implementation runs in quadratic time in the bit-length; see
also~\cite{Moller08} for discussion) computing
all the quotients explicitly, a continued fraction of
a $\Ot(k)$-bit rational number can be computed in $\Ot(k)$ time.
Needless to say, the quotients produced when running Euclid's algorithm have near-linear
total bit-length which
also means that the total bit-length of $a_0,\ldots,a_N$ is $\Ot(k)$~\cite{schonhage1971schnelle}.

Given a continued fraction, denote by $c_i$, $i=0,\ldots,N$, the \emph{$i$-th convergent}
\begin{equation*}
  c_i=a_0+\frac{1}{a_1+\frac{1}{\ddots+\frac{\vdots}{a_{i-1}+\frac{1}{a_i}}}}.
\end{equation*}
Each convergent $c_i$ is an irreducible fraction $\frac{p_i}{q_i}$, where $p_0,\ldots,p_N$ and $q_0,\ldots,q_N$
are given by $p_0=a_0$, $q_0=1$, $p_1=a_0a_1+1$, $q_1=a_1$ and for $i\geq 2$:
\begin{align*}
  p_{i}&=p_{i-1}\cdot a_i+p_{i-2},\\
  q_{i}&=q_{i-1}\cdot a_i+q_{i-2}.
\end{align*}
In other words, for $i\geq 2$ we have
\begin{equation}\label{eq:cmatrix}
  \begin{bmatrix}
    p_i & p_{i-1}\\
    q_i & q_{i-1}
  \end{bmatrix}=
  \begin{bmatrix}
    p_{i-1} & p_{i-2}\\
    q_{i-1} & q_{i-2}
  \end{bmatrix}\cdot
  \begin{bmatrix}
    a_i & 1\\
    1 & 0
  \end{bmatrix}=
  \begin{bmatrix}
    a_0 & 1\\
    1 & 0
  \end{bmatrix}\cdot
  \begin{bmatrix}
    a_1 & 1\\
    1 & 0
  \end{bmatrix}\cdot
  \ldots
  \cdot
  \begin{bmatrix}
    a_i & 1\\
    1 & 0
  \end{bmatrix}.
\end{equation}
\begin{lemma}{\upshape e.g., \cite[Section~1.1]{lovasz-alt}, \cite[Theorem~8]{bra}.}
  Let $b$ be a non-negative integer. Let $\ell\leq N$ be the largest integer such that $q_\ell< 2^b$.
  Let $t$ be the largest integer such that $tq_\ell+q_{\ell-1}< 2^b$.
  Then the sorted pair $\left\{\frac{p_\ell}{q_\ell},\frac{tp_\ell+p_{\ell-1}}{tq_\ell+q_{\ell-1}}\right\}$
  forms the best rational $b$-bit approximation of $\alpha$.
\end{lemma}
By the above lemma, given the continued fraction $[a_0;\ldots,a_N]$,
to compute the best rational approximation, we first need to find the index~$\ell$.
Since the denominators $q_1,\ldots,q_N$ are monotonically increasing,
$\ell$ can be found via binary search by testing $O(\log{k})$ numbers $q_j$ (recall that $N=\Ot(k)$)
and comparing them with $2^b$.
To implement a single step of binary search in $\Ot(k)$ time, it is enough to argue that a single pair $(p_j,q_j)$ can be computed in $\Ot(k)$ time.
This in turn can be done using the formula \eqref{eq:cmatrix}. One can evaluate the product
\begin{equation*}
   \begin{bmatrix}
    a_0 & 1\\
    1 & 0
  \end{bmatrix}\cdot
  \ldots
  \cdot
  \begin{bmatrix}
    a_j & 1\\
    1 & 0
  \end{bmatrix}
\end{equation*}
in $\Ot(|a_0|+\ldots+|a_j|)=\Ot(k)$ time, where $|a_i|$ denotes the bit-length of $a_i$, as follows.
Form a binary tree with the $j+1$ individual matrices in the leaves
and partial products of leaves in the corresponding subtrees of individual internal nodes.
Similarly as in Section~\ref{s:report-distance}, it is easy to see that the cost of evaluating the internal nodes at a single
level of this tree is $\Ot(|a_0|+\ldots+|a_j|)$.

Finally, given $q_\ell$ and $q_{\ell-1}$, the integer $t$ can be obtained in $\Ot(1)$ arithmetic operations.

\section{Proof of Lemma~\ref{l:bra-add}}\label{s:bra-add}

Suppose $\bra(\alpha,b)=\left(\frac{x_1}{y_1},\frac{x_2}{y_2}\right)$. Recall that $\alpha'=\frac{p}{q}$, where $0<q<2^{b'}$.

Let us define the following fractions:
\begin{itemize}
    \item $\frac{x'_1}{y'_1}$: the largest fraction with $y'_1 < 2^{b-b'}$ such that $\frac{x'_1}{y'_1} \leq \frac{x_1}{y_1} + \alpha'$;
    \item $\frac{x'_2}{y'_2}$: the smallest fraction with $y'_2 < 2^{b-b'}$ such that $\frac{x'_2}{y'_2} \geq \frac{x_2}{y_2} + \alpha'$.
\end{itemize}
Since $\frac{x_1}{y_1},\frac{x_2}{y_2}$ are $\Ot(b+\log(1+|\alpha|))$-bit rational numbers, one can compute both fractions $\frac{x'_1}{y'_1}$ and $\frac{x'_2}{y'_2}$
in $\Ot(b+\log(1+|\alpha|+|\alpha'|))$ time
by applying Lemma~\ref{l:bra} to
the respective fractions $\frac{x_1}{y_1} + \alpha'$ and $\frac{x_2}{y_2} + \alpha'$.
We will prove that $\left(\frac{x'_1}{y'_1},\frac{x'_2}{y'_2}\right)=\bra(\alpha+\alpha',b-b')$.

To this end, consider an arbitrary fraction $\frac{x}{y}$ with $y < 2^{b - b'}$. We will prove that 
$\frac{x}{y} \leq \frac{x'_1}{y'_1}$ if and only if $\frac{x}{y} \leq \alpha + \alpha'$.
An analogous property concerning the fraction
$\frac{x'_2}{y'_2}$ can be proved symmetrically.
The two properties imply that $\frac{x}{y}$ can only exist if $\frac{x}{y}=\alpha+\alpha'=\frac{x_1'}{y_1'}=\frac{x_2'}{y_2'}$.

The ``$\implies$'' implication follows immediately by the definition of $\frac{x'_1}{y'_1}$ and $\frac{x_1}{y_1} \leq \alpha$.
The ``$\impliedby$'' side does not hold only when $\frac{x'_1}{y'_1} < \frac{x}{y} \leq \alpha + \alpha'$.
Since $\frac{x}{y} > \frac{x'_1}{y'_1}$ and $y < 2^{b-b'}$, we have $\frac{x}{y} > \frac{x_1}{y_1} + \alpha'$ by the definition of $\frac{x'_1}{y'_1}$.
Hence, $\frac{x}{y} - \frac{p}{q} > \frac{x_1}{y_1}$. Equivalently, $\frac{xq' - yp}{yq} > \frac{x_1}{y_1}$. But we have $yq < 2^b$,
so by the definition of $\frac{x_1}{y_1}$ we have $\frac{x}{y} - \alpha' = \frac{xq - yp}{yq} > \alpha$.
Equivalently, $\frac{x}{y} > \alpha+\alpha'$, a contradiction.

\section{Proofs for discrete Exponential Start Time Clustering}
\label{s:proofs-estc}

The proofs in this section mirror their continuous analogs in \cite{MillerPVX15}.

\begin{proof}[Proof of \cref{lem:clustering-diameter}]
  For any $v \in V$, we have
  \[ \Pr\left[b_v \ge \frac{k \log n}{\alpha}\right] = \exp\left\{-\alpha \cdot \frac{k \log n}{\alpha}\right\} = \frac{1}{n^k}. \]
  Therefore, by the union bound:
  \[ \Pr\left[\max_{v \in V} b_v \ge \frac{k \log n}{\alpha} \right] \leq \sum_{v \in V} \Pr\left[ b_v \ge \frac{k \log n}{\alpha} \right] = \frac{1}{n^{k-1}}. \qedhere \]
\end{proof}

\begin{proof}[Proof of \cref{lem:edge-probab-clustering}]
  Fix an~edge $uv \in E$.
  Without loss of generality assume the graph is connected.
  For every vertex $x \in V$, define $a_x = \min(\dist_H(u, x), \dist_H(v, x))$ and $Y_x = b_x - a_x$.
  \begin{claim}
    If there exists $x \in V$ such that for all $y \in V \setminus \{x\}$ we have $Y_y \leq Y_x - 2$, then both $u$ and $v$ are in the cluster centered at $x$.
  \end{claim}
  \begin{proof}
    Note that for all $y \in V \setminus \{x\}$, we have that $\dist_H(u, x) \leq a_x + 1$ and $\dist_H(u, y) \geq a_y$.
    Therefore $\dist_H(u, y) - b_y > \dist_H(u, x) - b_x$.
    So $x$ is the unique vertex minimizing the value of $\dist_H(u, x) - b_x$ and $u$ must have been assigned to the cluster centered at $x$.
    The same reasoning applies to $v$.
  \end{proof}
  Hence, the probability of both $u$ and $v$ being in the same cluster is lower-bounded by the probability that the maximum value $Y_x$ for $x \in V$ is at least $2$ larger than the second maximum value $Y_y$ for $y \in V$.
  
  Introduce an~arbitrary linear order $<$ on $V$ and define a~linear order $<$ on $\mathbb{N} \times V$ as follows: $(a_1, v_1) < (a_2, v_2)$ if either $a_1 < a_2$, or $a_1 = a_2$ and $v_1 < v_2$.
  Consider the following event $E_{x, y, Y}$: among the set of pairs $(Y_v, v)$ for $v \in V$, the maximum pair is $(Y_x, x)$, the second maximum pair is $Y_y$, and $Y = Y_y$.
  Note that the events $E_{x, y, Y}$ partition the probability space, so by the law of total probability it is enough to show that, for each event $E_{x, y, Y}$ with non-zero probability,
  \[ \Pr[Y_x \geq Y + 2 \,\mid\, E_{x,y,Y}] \geq e^{-2\alpha}. \]

  Fix an~event $E_{x,y,Y}$ with non-zero probability.
  Conditional on this event, each variable $Y_v$ for $v \in V$ has its range restricted: we have $(Y_x, x) > (Y, y)$, i.e., $Y_x > Y$ if $x < y$ and $Y_x \geq Y$ otherwise; and $(Y_v, v) < (Y, y)$ for $v \notin \{x, y\}$, i.e., $Y_v \le Y$ if $v < y$ and $Y_v < Y$ otherwise.
  Note, however, that even with these restrictions, all variables $Y_v$ for $v \in V$ are independent, so we have
  \[ \Pr[Y_x \geq Y + 2 \,\mid\, E_{x,y,Y}] = \Pr[Y_x \geq Y + 2 \,\mid\, (Y_x, x) > (Y_y, y)]. \]
  Since $Y_x \geq Y_y + 2$ implies $(Y_x, x) > (Y_y, y)$, which in turn implies $Y_x \geq Y_y$, we also get that
  \[ \Pr[Y_x \geq Y + 2 \,\mid\, E_{x,y,Y}] \geq \Pr[Y_x \geq Y + 2 \,\mid\, Y_x \geq Y]. \]
  Then we have that
  \begin{align*}
  \Pr[Y_x \geq Y_y + 2 \,\mid\, Y_x \geq Y_y] &= \Pr[b_x \geq Y + a_x + 2 \,\mid\, b_x \geq Y + a_x] \\
    &= \Pr[b_x \geq Y + a_x + 2 \,\mid\, b_x \geq \max(0, Y + a_x)] \\
    &\stackrel{(\star)}{=} \Pr[b_x \geq Y + a_x + 2 - \max(0, Y + a_x)] \\
    &\stackrel{(\square)}{\geq} \Pr[b_x \geq 2] = e^{-2\alpha},
  \end{align*}
  where in $(\star)$, we use the memoryless property of the geometric distribution, and in $(\square)$, we use the fact that $(Y + a_x) + 2 - \max(0, Y + a_x) \leq 2$.
\end{proof}

\section{Exponentially small differences of sums of short rationals} \label{s:small-diff}

In this Appendix, we are going to show the following lemma and discuss its consequences.

\begin{lemma}
	Let $n$ be a positive integer. 
	Then, there exist rational numbers $p_1, \ldots, p_n, q_1, \ldots, q_n \in \Q[n]$ such that the difference $(p_1 + \ldots p_n) - (q_1+\ldots+q_n)$ is positive, but belongs to $2^{\Oh(-\frac{n}{\log n})}$.
\end{lemma}
\begin{proof}
	Let $P$ be the set of prime numbers that are smaller than $n$. For every $S \subseteq P$ let us define $r_S = \sum_{s \in S} \frac{1}{s}$.
	
	We would like to prove that all values $r_S$ are different. In order to do this, it suffices to prove the following claim:
	\begin{claim}
		Let $S \subseteq P$ and $S = \{s_1, \ldots, s_c\}$. Then, if we express $r_S$ in the irreducible form, its denominator is equal to $s_1 \cdot  \ldots \cdot s_c$.
	\end{claim}
	\begin{proof}
		We recall a simple fact that in an irreducible form of a sum of rational numbers, the denominator divides the product of denominators of numbers in the sum.
		
		Let us denote the irreducible form of $\frac{1}{s_2} + \ldots + \frac{1}{s_c}$ as $\frac{a}{b}$. Note that $s_1$ does not divide $b$. Then, $r_S = \frac{1}{s_1} + \frac{a}{b} = \frac{as_1 + b}{s_1b}$. Since $s_1$ does not divide $v$, it does not divide $as_1 + b$. Hence, it divides the denominator of $r_S$ in the irreducible form (recall that $s_1$ is prime). Similarly, any other $s_i$ divides it as well. As all $s_1, \ldots, s_c$ are different prime numbers, the denominator of $r_S$ has to be divisible by their product. This finishes the proof.
	\end{proof}

	We conclude the all the numbers of form $r_S$ are different and contained in the interval $[0, n]$. Hence, some two of them have to differ by at most $\frac{n}{2^{|P|}-1}$. However, as the prime number theorem states, $|P|=\Theta(\frac{n}{\log n})$. Recall that each $r_S$ is a sum of at most $n$ rational numbers from $\Q[n]$ (note that we are allowed to set some numbers $p_i$, $q_i$ to zeros). 
\end{proof}

For a positive integer $n$, let us now create two vertices, call them $s$ and $t$, and two paths of $n$ edges from $s$ to $t$. Edges on the first path should have weights $p_1, \ldots, p_n$ and on the second path $q_1, \ldots, q_n$. Their weights will differ by $2^{-\widetilde{\Omega}(n)}$, as claimed in the overview.

Let us point out that a priori, one can expect that a difference of such sums can be as low as $2^{-\Theta(n \log n)}$ (as it can be expressed as a rational with the denominator at most $n^{2n}$), while we have shown the construction for a weaker bound of $2^{-\Theta(\frac{n}{\log n})}$ only. However, this is optimal up to polylogarithmic factors in the exponent, so we cannot improve the naive bound on the lengths of bit representations of sums of short rationals by more than a polylogarithmic factor, emphasizing the difficulty of dealing with rational arithmetic.